\documentclass[a4paper,fleqn,usenatbib]{mnras}
\pdfoutput=1
\makeatletter
\def\maxwidth{ %
  \ifdim\Gin@nat@width>\linewidth
    \linewidth
  \else
    \Gin@nat@width
  \fi
}
\makeatother

\usepackage[]{graphicx}
\usepackage[]{color}

\definecolor{fgcolor}{rgb}{0.345, 0.345, 0.345}

\usepackage{framed}
\makeatletter
 {\par\unskip\endMakeFramed%
 \at@end@of@kframe}
\makeatother

\definecolor{shadecolor}{rgb}{.97, .97, .97}
\definecolor{messagecolor}{rgb}{0, 0, 0}
\definecolor{warningcolor}{rgb}{1, 0, 1}
\definecolor{errorcolor}{rgb}{1, 0, 0}
\newenvironment{knitrout}{}{} 

\usepackage{alltt}

\usepackage{mathptmx}

\usepackage[T1]{fontenc}
\usepackage{ae,aecompl}

\usepackage{aas_macros}
\usepackage{amsmath}
\usepackage{amssymb}
\usepackage{longtable}
\usepackage{mathptmx}
\usepackage{multirow}
\usepackage{url}
\usepackage[section]{placeins}
\usepackage[normalem]{ulem}
\usepackage{wasysym}

\newcommand{\dg}{^{\circ}}

\title[Optical linear polarization of 74 white dwarfs]{Optical linear polarization of 74 white dwarfs with the RoboPol polarimeter}

\author[M.~\.Zejmo et al.]{Micha\l{}~\.Zejmo$^1$, 
Aga S\l{}owikowska$^1$,
Krzysztof~Krzeszowski$^1$, 
Pablo Reig$^{2,3}$, 
\newauthor
and Dmitry Blinov$^{2, 3,4}$\\
$^1$ Janusz Gil Institute of Astronomy, University of Zielona G\'ora, Lubuska 2,
65-265 Zielona G\'ora, Poland\\
$^2$ Foundation for Research and Technology, 71110 Heraklion, Crete, Greece\\
$^3$ University of Crete, Physics Department, PO Box 2208, 710 03 Heraklion, Crete, Greece\\
$^4$ Astronomical Institute, St. Petersburg State University,Universitetsky pr. 28, Petrodvoretz, 198504 St. Petersburg,
Russia \\
}

\date{Accepted XXX. Received YYY; in original form ZZZ}
\pubyear{2016}
\IfFileExists{upquote.sty}{\usepackage{upquote}}{}
\begin{document}
\label{firstpage}
\pagerange{\pageref{firstpage}--\pageref{lastpage}} \pubyear{2016}

\maketitle

\begin{abstract}
We present the first linear polarimetric survey of white dwarfs (WDs).
Our sample consists of WDs of DA and DC spectral
types in the SDSS \textit{r} magnitude range from
13 to 17. 
We performed polarimetric observations with the RoboPol polarimeter 
attached to the 1.3-m telescope at the Skinakas Observatory.
We have 74 WDs in our sample, of which almost all
are low polarized WDs with polarization degree (PD) smaller
than 1\%, while only~2 have PD higher than 1\%.
There is an evidence that on average the isolated WDs of DC type have higher PD (with median PD of 0.78\%) 
than the isolated DA type WDs (with median PD of 0.36\%). On the other hand, the median PD  
of isolated DA type WDs is almost the same, i.e. 0.36\%
as the median PD of DA type white dwarfs in binary systems with red dwarfs (dM type), i.e. 0.33\%. 
This shows, as expected, that there is no contribution to the PD from the companion
if the WD companion is the red dwarf, which is the most common situation
for WDs binary systems.
We do not find differences in the polarization degree between magnetic and non-magnetic WDs.
Because  97\%
of WDs in our sample have PD lower than 1\%,  they can be used as faint zero--polarized standard star in the magnitude range from 13 up to 17 of SDSS \textit{r} filter. 
They cover the Northern sky between 13 hour to 23 hour in right ascension 
and from \ensuremath{-11}$\dg$ to 
78$\dg$ in declination.
Additionally, we found that for low extinction values (< 0.04) the best
model that describes the dependence of PD on E(B--V) is given by the equation:
$\rm{PD_{max, ISM}}[\%] = 0.65~\rm{E(B-V)}^{0.12}$.
\end{abstract}


\section{Introduction}
\label{sec:introduction}

In the last years the interest in optical polarimetry has grown significantly \citep[e.g.][]{Marscher2010}. The reason for this boom is that polarimetric measurements give an invaluable additional constrain on theoretical models that neither the photometry, astrometry nor spectrometry can provide. These studies include all kinds of astrophysical objects. There are regular monitoring campaigns to study the polarization
changes of AGNs in the optical domain, as for example the optical monitoring
of selected blazars with the
RoboPol polarimeter \citep{Pavlidou2014}. There are also 
optical polarization studies of isolated neutron stars
including pulsars \citep[e.g.][]{Slowikowska2009, Lundqvist2011,
Moran2013, Moran2014, Mignani2015}
and magnetars \citep[e.g.][]{Wang2015}, as well as neutron
stars in high mass X-ray binaries 
\citep[e.g.][S\l{}owikowska et al. in preparation]{Reig2014}
and low mass X-ray binaries \citep[e.g.][]{Baglio2014},
not to mention polarization studies of GRBs \citep{Mundell2013}
and of polarized light from exoplanets
for which a dedicated detector, i.e. Spectro-Polarimetric
High-contrast Exoplanet REsearch (SPHERE) at VLT has been recently built\footnote{\url{https://www.eso.org/sci/facilities/paranal/instruments/sphere.html}}.

The scientific community has started to use
polarimetric measurements extensively to study stellar and non-stellar
objects. However, reaching fainter objects by using
infrastructure with larger mirror introduced
a serious problem, namely, the lack of faint
polarization standards of both types --- the zero-polarized and polarized ones.
Each measurement using a polarimeter or spectropolarimeter needs to be properly calibrated. Thus, the polarized standards are necessary to establish
the intrinsic depolarization caused by the instrument, while
the zero-polarized standards are necessary to get the instrumental polarization
(e.g. the RINGO3 polarmeter at the Liverpool Telescope, see \citet{Slowikowska2016}).

The aim of our work is twofold: {\em i)} to perform a statistical analysis of the linear polarization properties of white dwarfs sample and {\em ii)} to provide observers with new faint linear polarimetric standard sources.

There are more than 23,000 WDs known up to date (see Sec.~\ref{sec:sample}). For many of them their spectral type is known. Most of of WDs atmospheres are hydrogen-rich atmospheres (DA), while almost all the rest are helium-rich (DB). However, a significant
fraction of WDs also contain trace elements in their atmospheres and therefore they
are labelled with Z for metals or Q for carbon, as for example DZ or DQ. There are also cases when the WD spectrum does not show any strong lines, but still their
atmospheres are helium-rich. Such WDs are classified as DC type WDs. 
WDs with magnetic fields stronger than 1MG can be detected via Zeeman splitted
lines, while weaker magnetic fields can be detected using spectropolarimetry.
However, because DC type WDs spectra do not have strong spectral lines, therefore it is not possible to use the Zeeman effect to measure their magnetic field strength. Previous studies showed that around 10\% up to 20\% of all WDs are magnetised with strong magnetic field \citep[][and references therein]{Ferrario2015}. The PD of magnetised WDs is between a fraction of a percent up to a few percent, as for example in case of GD~229 that has almost $8\%$ of linear PD in R band \citep{Berdyugin1999}. The population of magnetised WDs can be even larger because the magnetic field of many sources is unknown. 

Linear polarimetric population study can help to select WDs as good candidates for stable polarimetric calibration sources. Moreover, for each WD we have information about whether it is an isolated WD or WD in a binary system. In most cases the companion is a low mass red dwarf. Magnetised WDs in binary systems with low mass star that are in contact are classified as the magnetic cataclysmic variables (MCVs,
i.e. polars and intermediate polars) and they represent one fourth of the whole CVs population \citep{Wickramasinghe2000}.
There are also binary systems composed of WD and low mass star that are close but are not in contact, i.e. pre-CVs, however none of such systems with magnetic WD is known so far \citep{Liebert2015}. There are close double degenerate systems and common proper motion binaries as well. This allows us to study the dependence of WD polarimetric properties on singularity or binarity, taking into account the type of the binary.

White dwarfs are commonly used as zero polarization standard stars.
In the literature we can find eleven white dwarfs used for this purpose.
Two of them, i.e. G191-B2B (PD=0.09\%) and GD319 (PD=0.045\%), were
proposed by \cite{Turnshek1990} as the HST polarimetric standards,
whereas another nine were proposed by \cite{Fossati2007}
as main zero polarization standards for the FORS1 instrument on the VLT.
The main goal of \cite{Fossati2007} was to find a group of faint
polarized and non-polarized standard stars that can be used for calibration of big telescopes.
His sample consists of 30 stars of different types in the magnitude range from 6 to 14. 
However, WDs given by \cite{Fossati2007} are in the magnitude range from 11 to 13, with only one exception of 14 mag.
In our work we propose to extend existing standard lists with
additional 74 WDs as low linear polarization standards. The biggest advantage of WDs from our sample is that they are even fainter,
i.e. in the SDSS $r$ magnitude range from 13.2 
(WD2149+021) up to 17 
(WD2213+317), than those already available in the literature. In this way our sample is complementary
to the earlier work. A larger group of zero polarization standards allows to find
visible standard in convenient time of the night and position on the sky.

There were many polarization studies of large samples of white dwarfs
conducted so far, for example: \citet{Angel1981}; 
\citet{Schmidt1995}; \citet{Putney1997}; \citet{Kawka2007}; 
\citet{Jordan2007}; \citet{Kawka2012}; 
\citet{Landstreet2015}.
Recently, \citet{Bagnulo2015} published a spectropolarimetric 
catalogue of 809 objects, obtained with the FORS/VLT instrument, that includes 70 WDs.
However, there is only one WD common in both lists, ours and \citet{Bagnulo2015}, i.e. WD2149+021.
The crucial difference between above mentioned studies and our work is 
that our observations are the first WDs linear polarization survey, whereas 
the others measured the circular polarization. 

We describe the selection method in Sec.~\ref{sec:sample}, the observations in 
Sec.~\ref{sec:obs}, while the data analysis in Sec.~\ref{sec:data}. Results and 
conclusions are given in Sec.~\ref{sec:results} and Sec.~\ref{sec:summary}, respectively. 

\section{Selected sample}
\label{sec:sample}

Our sample was built up from the following catalogues: the White Dwarf Catalogue of Villanova University\footnote{\url{http://www.astronomy.villanova.edu/WDcatalog/}} (\cite{McCook1999}, hereafter VUWDC), ''SDSS DR7 White Dwarf Catalog" (\cite{Kleinman2013}, here-after DR7WDC) and ''Post-common envelope binaries from SDSS-XIV. The DR7 white dwarf-main-sequence binary catalogue" (\cite{Rebassa2012}, hereafter DR7WDC-bin). We gathered 23,068 objects from those catalogues. However, it should be mentioned that there are 14,235, 18,913 and 2,248 WDs in the VUWDC, DR7WDC and DR7WDC-bin, respectively. Some of WDs are repeated in all three catalogues, therefore the final number of WDs, i.e. 23,068, is smaller than just a simple sum of WDs from these catalogues. Once the sky coordinates were extracted, we searched for their photometric brightness measurements in the following sky surveys: SDSS \citep{York2000, Ahn2012}, UKIDSS (\cite{Lawrence2007}, \cite{Casali2007}, \cite{Hewett2006}, \cite{Hodgkin2009}, \cite{Hambly2008}), 2MASS \citep{Skrutskie2006} and WISE \citep{Wright2010}. 
Of the 23,068 WDs, 18,877 have known spectral type. Based
on the previously mentioned catalogues, the distribution of different types of WDs is as follows: DA - strong hydrogen lines 84.2\%,  DB - strong He I lines 8.2\%, DC - no strong lines, continuous spectrum 4\%, DO - strong He II lines 0.3\%, DQ - strong carbon lines 1.6\%, DZ - strong metal lines, excluding carbon 1.6\%. The contribution of the binary systems in the WD population is on the level of 13\%. 
However, this number can be higher in reality because the catalogues only indicate the confirmed binary systems.

Due to observational constraints, we selected objects brighter than 17 mag and visible from the 1.3-m telescope at the Skinakas Observatory\footnote{\url{http://skinakas.physics.uoc.gr/en/}} in Greece during the time span when the observatory operates, i.e. from April to November.
This selection resulted in 618 WDs of DA spectral type, 54 WDs of DB spectral type, 15 WDs of DC spectral type, and 70 with other or unknown spectral classification.  There is no bias in the number of WDs in our Skinakas sample due to the fact that we only impose the brightness limit on SDSS $r$ < 17 mag and apply visibility cuts. The sample is not biased with respect to binarity either as $\sim$20\% of the selected objects are confirmed binaries. We conclude that our sample is a complete unbiased sample of the whole population.

\section{Observations}
\label{sec:obs}

Our observations were performed at the Skinakas Observatory  located in Crete, Greece. For our study we used the RoboPol\footnote{\url{http://robopol.org/}} polarimeter that is a linear polarimeter with two Wollaston prisms and two half-wave plates. It allows to measure the Stokes I, Q, U parameters simultaneously for all stellar objects within 13 $\times$ 13 arcmin$^2$ field of view and the scale of 0.435"/pixel \citep{King2014}. The measurements were performed with an aperture defined as 2.5 $\times$ FWHM, where FWHM is an average full width at half maximum of stellar images, which has a median value of 1.7 arcsec (4.0 pixels). All observations with the RoboPol are performed with the R Johnston filter. The exposure time was adjusted according to the brightness of each target, which was estimated during the short pointing exposures. It was calculated in such a way to ensure S/N ratio equal to 10 in the PD for a 2.5\% polarized source. Moreover, 300 seconds overhead was needed for each target as operational time that includes slew time and positioning time. 

We performed observations of 74 WDs between May 20th, 2014 and June 6th, 2014 with one single observation on October 24th, 2013, collecting a total observing time of around 14 hours. While performing the observations, we firstly concentrated on the DA and DC type WDs. Because the observations were cut short due to bad weather conditions, we were not able to observe any of the selected DB WDs. Each WDs was observed
only once, but single observation consists of a few exposures with appropriate
calculated exposure time. Later, all the images were aligned with subpixel accuracy and stacked together. The observing log is presented in Tab.~\ref{tab:log}, where the WD name, coordinates, binarity indication, spectral type, brightness
and distance along with their corresponding errors, E(B-V), as well as the observation date are given. All \textit{r} brightness values were obtained from the SDSS data base \citep{Ahn2012}. Their corresponding errors are not greater than 0.01 mag.
At first, distances
were taken from the spectroscopic measurements published by \cite{Holberg2008},
because they also provided the errors. In other cases the most recent distance measurements available in the literature were taken. 
We collected 51 distance measurements and 7 distance errors. All references for the spectral type and distance are provided in the table header. The values of E(B-V) were obtained from \cite{Schlafly2011} using the NASA/IPAC Extragalactic Database \citep{NED2009}.

There are 53 isolated WDs and 
21 WDs in binary systems in our sample.
Isolated WDs group consists of 49 DA
WDs and 4 DC WDs (WD1402+649, WD1425+495, WD1524+566, WD1712+215), 
whereas the group of WDs in binary systems 
consists of 14 systems where the companion of the DA WD
is the main sequence star, mainly red dwarf (11 DA+dM systems, as well as 
2 DA+K7, and 1 DA+F8), 
and 7 double degenerated systems (DDSs) out of which 3 are classified as
the DA+DA binaries, 2 are classified as DA+DC and the last 2 are WD+WD and WD+sdB.
There are two magnetic WDs in our sample, i.e. WD1639+537 (GD 356) and WD1658+440 (PG 1658+440).

The RoboPol sensitivity is optimized for a source at the centre of the field by using a cross--shaped mask in the telescope focal plane. 
It reduces the sky background by a factor of 4 comparing the central
target to the field stars. However, the mask also obscures part of the field,
reducing the effective field of view. This plays the biggest role in case of
one of our sources, i.e. WD1258+593 (PG1258+593, DA+DAH, $m_r = 15.5$).
This source is a common proper motion binary system with the 
the separation between components of 16 arc seconds
\citep{Girven2010} and its magnetic companion was behind the mask during the observations.  Therefore, we were not able to measure its polarization.

The SDSS \textit{r} brightness distribution of the observed
WDs is shown in Fig.~\ref{fig:histWD} with the mean of 15.3, whereas their distribution over the sky in the equatorial coordinates $(\alpha, \delta)$ as well as in the galactic coordinates $(l, b)$ are shown in the left and right panels of Fig.~\ref{fig:histWD2}, respectively. There is a clear lack of WDs in our sample located in the Galactic
latitude range between -20$\dg$ and +20$\dg$, i.e. in the location of the Galactic disk.
It is caused by the fact, that surveys very often avoid this region of the sky.

\begin{table*}
\centering
\caption{RoboPol observation log of 74 white dwarfs: WD name, coordinates, binarity indication, where Y stands for WD in binary system and N for an isolated WD, WD spectral type, WD spectral type reference, brightness in SDSS \textit{r} filter, distance with error and reference, E(B-V) and observation date.
Spectral type references: B92 - \citet{Berg1992}, F05 - \citet{Farihi2005}, F97 - \citet{Ferrario1997}, G10 - \citet{Girven2010}, H12 - \citet{Holberg2012}, K13 - \citet{Kleinman2013}, K09 - \citet{Koester2009}, L13- \citet{Limoges2013}, M95 - \citet{Marsh1995}, M99 - \citet{McCook1999}, M05 - \citet{Morales-Rueda2005}, S10 - \citet{Schreiber2010}, S05 - \citet{Silvestri2005}. Distance references: G11 - \citet{Gianninas2011}, H08 - \citet{Holberg2008}, S14 - \citet{Sion2014}. WDs common with  \citet{Schmidt1995} and \citet{Putney1997} 
 are marked with $^\dagger$.} 
\label{tab:log}
\scalebox{0.8}{
\begin{tabular}{lccccccccccc}
  \hline
WD Name & RA & Dec & Binary & Type & Type Ref. & r & Dist. & $\sigma_\mathrm{Dist}$ & Dist. Ref. & E(B-V) & Obs. Date \\ 
  & [hh:mm:ss] & [dd:mm:ss] & & & & [mag] & [pc] & [pc] & \\ \hline
WD1257+037$^\dagger$ & 13:00:09.06 & +03:28:41.0 & N & DA & M99 & 15.6 &  &  &  & 0.02 & 2014-05-22 \\ 
  WD1258+593 & 13:00:35.19 & +59:04:15.5 & Y & DA+DAH & G10 & 15.5 & 65.0 &  & G11 & 0.01 & 2014-05-22 \\ 
  WD1259+674 & 13:01:21.10 & +67:13:07.3 & N & DA & M99 & 16.4 &  &  &  & 0.01 & 2014-05-22 \\ 
  WD1310+583$^\dagger$ & 13:12:57.89 & +58:05:11.2 & N & DA & M99 & 14.2 & 21.1 &  & S14 & 0.01 & 2014-06-05 \\ 
  WD1317+453$^\dagger$ & 13:19:13.71 & +45:05:09.9 & Y & WD+WD & M95 & 14.2 & 49.0 &  & G11 & 0.03 & 2014-05-31 \\ 
  WD1319+466$^\dagger$ & 13:21:15.08 & +46:23:23.7 & N & DA & M99 & 14.7 & 36.0 &  & G11 & 0.01 & 2014-06-05 \\ 
  WD1334+070 & 13:36:33.68 & +06:46:26.8 & Y & DA+DA & K09 & 15.5 &  &  &  & 0.03 & 2014-06-05 \\ 
  WD1344+106 & 13:47:24.36 & +10:21:37.9 & N & DA & M99 & 15.1 & 20.0 & 1.4 & H08 & 0.03 & 2014-06-05 \\ 
  WD1344+572$^\dagger$ & 13:46:02.07 & +57:00:32.7 & N & DA & M99 & 13.6 & 20.0 &  & S14 & 0.00 & 2014-05-31 \\ 
  WD1349+144 & 13:51:53.92 & +14:09:45.4 & Y & DA+DA & K09 & 15.4 & 85.0 &  & G11 & 0.02 & 2014-06-05 \\ 
  WD1401+005 & 14:03:45.31 & +00:21:36.0 & N & DA & M99 & 15.0 & 463.0 &  & G11 & 0.03 & 2014-06-05 \\ 
  WD1402+649 & 14:03:53.49 & +64:39:53.7 & N & DC & M99 & 16.9 &  &  &  & 0.01 & 2014-05-31 \\ 
  WD1407+374 & 14:09:23.26 & +37:10:48.8 & Y & DA+dM0 & S10 & 14.8 &  &  &  & 0.01 & 2014-06-05 \\ 
  WD1407+425$^\dagger$ & 14:09:45.25 & +42:16:00.8 & N & DA & M99 & 15.1 & 31.0 &  & G11 & 0.01 & 2014-05-31 \\ 
  WD1408+323$^\dagger$ & 14:10:26.95 & +32:08:36.1 & N & DA & M99 & 14.2 & 39.5 & 4.4 & H08 & 0.01 & 2014-06-05 \\ 
  WD1415+132 & 14:17:40.25 & +13:01:48.7 & Y & DA+dM & K09 & 15.5 & 208.0 &  & G11 & 0.02 & 2014-06-05 \\ 
  WD1420+518 & 14:22:41.92 & +51:35:37.9 & Y & DA+K7 & K13 & 15.7 &  &  &  & 0.01 & 2014-06-05 \\ 
  WD1421+318 & 14:23:40.72 & +31:34:59.8 & N & DA & M99 & 15.6 & 112.0 &  & G11 & 0.02 & 2014-06-05 \\ 
  WD1422+497 & 14:24:40.53 & +49:29:58.1 & Y & DA+K7 & K13 & 15.7 &  &  &  & 0.02 & 2014-06-05 \\ 
  WD1425+495 & 14:26:59.41 & +49:21:00.4 & N & DC & M99 & 16.8 &  &  &  & 0.02 & 2014-05-31 \\ 
  WD1428+373 & 14:30:42.62 & +37:10:15.4 & Y & DA+DC & F05 & 15.6 & 97.0 &  & G11 & 0.01 & 2014-06-05 \\ 
  WD1429+373 & 14:31:56.67 & +37:06:30.0 & N & DA & M99 & 15.5 & 123.0 &  & G11 & 0.01 & 2014-05-31 \\ 
  WD1434+328 & 14:36:49.60 & +32:37:34.5 & Y & DA+dM0e & K13 & 15.9 &  &  &  & 0.01 & 2014-06-05 \\ 
  WD1440$-$025 & 14:43:35.06 & -02:43:52.1 & Y & DA+dMe & B92 & 15.7 &  &  &  & 0.06 & 2014-06-05 \\ 
  WD1446+286$^\dagger$ & 14:48:14.09 & +28:25:11.8 & N & DA & M99 & 14.8 & 48.0 &  & G11 & 0.02 & 2014-06-05 \\ 
  WD1447+049 & 14:50:09.84 & +04:41:45.7 & N & DA & M99 & 15.6 &  &  &  & 0.04 & 2014-06-05 \\ 
  WD1449+168 & 14:52:11.40 & +16:38:03.5 & Y & DA+dM3 & F05 & 15.6 & 102.0 &  & G11 & 0.02 & 2014-06-05 \\ 
  WD1503$-$093 & 15:06:19.43 & -09:30:20.8 & N & DA & M99 & 15.3 & 57.0 &  & G11 & 0.08 & 2014-06-05 \\ 
  WD1507$-$105 & 15:10:29.04 & -10:45:19.4 & N & DA & M99 & 15.4 & 51.0 &  & G11 & 0.09 & 2014-06-05 \\ 
  WD1515+668 & 15:15:52.89 & +66:42:42.8 & N & DA & M99 & 15.5 & 32.0 &  & G11 & 0.02 & 2014-05-31 \\ 
  WD1518$-$003 & 15:21:30.84 & -00:30:55.7 & N & DA & M99 & 15.5 &  &  &  & 0.06 & 2014-05-31 \\ 
  WD1524+566 & 15:25:42.95 & +56:29:08.8 & N & DC & M99 & 16.7 &  &  &  & 0.01 & 2014-05-31 \\ 
  WD1525+257 & 15:27:36.50 & +25:35:06.0 & N & DA & M99 & 15.9 & 87.0 &  & G11 & 0.04 & 2014-06-05 \\ 
  WD1527+090 & 15:29:50.41 & +08:55:46.4 & N & DA & M99 & 14.5 & 53.0 &  & G11 & 0.03 & 2014-05-31 \\ 
  WD1534+503 & 15:36:15.83 & +50:13:51.1 & N & DA & M99 & 15.8 & 38.0 &  & G11 & 0.01 & 2014-06-01 \\ 
  WD1538+269 & 15:40:23.45 & +26:48:29.8 & Y & WD+sdB & M05 & 14.1 &  &  &  & 0.04 & 2014-05-20 \\ 
  WD1538+333 & 15:40:33.38 & +33:08:52.5 & N & DA & M99 & 15.1 & 22.7 &  & S14 & 0.03 & 2014-06-01 \\ 
  WD1548+149 & 15:51:15.43 & +14:46:59.2 & N & DA & M99 & 15.3 & 83.0 &  & G11 & 0.04 & 2014-06-01 \\ 
  WD1553+353$^\dagger$ & 15:55:02.10 & +35:13:24.0 & N & DA & M99 & 15.0 &  &  &  & 0.02 & 2014-05-31 \\ 
  WD1601+581$^\dagger$ & 16:02:41.70 & +57:58:13.0 & N & DA & M99 & 14.2 & 42.0 &  & G11 & 0.02 & 2014-05-31 \\ 
  WD1606+422$^\dagger$ & 16:08:22.20 & +42:05:43.2 & N & DA & M99 & 14.0 & 45.0 & 7.1 & H08 & 0.01 & 2014-05-31 \\ 
  WD1610+166 & 16:13:02.31 & +16:31:55.5 & N & DA & M99 & 15.9 & 65.4 & 18.4 & H08 & 0.05 & 2014-06-01 \\ 
  WD1614+160 & 16:17:08.76 & +15:54:38.5 & N & DA & M99 & 15.7 &  &  &  & 0.03 & 2014-06-05 \\ 
  WD1630+089 & 16:32:33.18 & +08:51:22.6 & N & DA & M99 & 14.9 & 13.2 &  & S14 & 0.05 & 2014-05-31 \\ 
  WD1631+396 & 16:33:39.31 & +39:30:53.5 & N & DA & M99 & 14.7 & 51.0 &  & G11 & 0.01 & 2014-05-31 \\ 
  WD1632+177$^\dagger$ & 16:34:41.50 & +17:36:32.0 & N & DA & M99 & 13.4 & 15.0 &  & G11 & 0.05 & 2014-05-31 \\ 
  WD1636+160 & 16:38:40.40 & +15:54:17.0 & N & DA & M99 & 15.8 & 68.0 &  & G11 & 0.06 & 2014-06-05 \\ 
  WD1636+351 & 16:38:26.32 & +35:00:11.9 & N & DA & M99 & 15.1 & 111.0 &  & G11 & 0.02 & 2014-05-31 \\ 
  WD1637+335$^\dagger$ & 16:39:27.82 & +33:25:22.3 & N & DA & M99 & 14.7 & 28.6 & 2.6 & H08 & 0.02 & 2014-05-31 \\ 
  WD1639+537$^\dagger$ & 16:40:57.16 & +53:41:09.6 & N & DAP & F97 & 15.0 & 21.1 &  & S14 & 0.02 & 2014-05-31 \\ 
  WD1641+387$^\dagger$ & 16:43:02.29 & +38:41:17.3 & N & DA & M99 & 14.8 & 43.0 &  & G11 & 0.01 & 2014-05-31 \\ 
  WD1643+143 & 16:45:39.14 & +14:17:46.2 & Y & DA+dM & K09 & 15.1 & 133.0 &  & G11 & 0.07 & 2014-05-31 \\ 
  WD1647+375$^\dagger$ & 16:49:20.29 & +37:28:21.2 & N & DA & M99 & 15.2 & 82.0 &  & G11 & 0.02 & 2014-05-31 \\ 
  WD1654+637 & 16:54:29.01 & +63:39:21.0 & N & DA & M99 & 15.9 & 89.0 &  & G11 & 0.02 & 2014-06-06 \\ 
  WD1655+215$^\dagger$ & 16:57:09.85 & +21:26:50.2 & N & DA & M99 & 14.4 & 23.3 & 1.7 & H08 & 0.05 & 2014-05-31 \\ 
  WD1658+440$^\dagger$ & 16:59:48.41 & +44:01:04.5 & N & DAP & M99 & 15.0 & 22.0 &  & S14 & 0.01 & 2014-05-31 \\ 
  WD1659+303$^\dagger$ & 17:01:08.02 & +30:15:35.8 & Y & DA+dM2 & F05 & 15.1 & 50.0 &  & G11 & 0.03 & 2014-05-31 \\ 
  WD1706+332 & 17:08:52.06 & +33:12:58.5 & Y & DA+F8 & H12 & 16.1 & 81.0 &  & G11 & 0.03 & 2014-05-31 \\ 
  WD1712+215 & 17:14:30.50 & +21:27:11.3 & N & DC & M99 & 16.6 &  &  &  & 0.05 & 2014-05-31 \\ 
  WD1713+695$^\dagger$ & 17:13:06.12 & +69:31:25.7 & N & DA & M99 & 13.5 & 27.0 &  & G11 & 0.03 & 2014-05-31 \\ 
  WD1723+563 & 17:24:06.14 & +56:20:03.1 & Y & DA+dMe & K13 & 16.4 &  &  &  & 0.02 & 2014-06-01 \\ 
  WD1734+575 & 17:35:13.28 & +57:30:12.2 & N & DA & M99 & 16.7 &  &  &  & 0.05 & 2014-06-06 \\ 
  WD1738+669 & 17:38:02.53 & +66:53:47.8 & N & DA & M99 & 14.8 & 147.0 &  & G11 & 0.04 & 2014-06-06 \\ 
  WD1827+778 & 18:25:09.28 & +77:55:35.1 & N & DA & M99 & 16.3 & 393.0 &  & G11 & 0.07 & 2014-06-06 \\ 
  WD1833+644 & 18:33:29.21 & +64:31:52.1 & Y & DA+dM2e & K13 & 16.4 & 376.0 &  & G11 & 0.04 & 2014-06-06 \\ 
  WD1842+412 & 18:44:12.63 & +41:20:29.4 & Y & DA+dM6e & K13 & 16.4 &  &  &  & 0.07 & 2014-06-01 \\ 
  WD2006+615 & 20:06:54.88 & +61:43:10.4 & N & DA & L13 & 16.3 &  &  &  & 0.14 & 2014-06-01 \\ 
  WD2058+083 & 21:01:13.37 & +08:35:09.4 & N & DA & M99 & 14.1 & 134.0 &  & G11 & 0.07 & 2014-06-01 \\ 
  WD2126+734$^\dagger$ & 21:26:57.66 & +73:38:44.5 & Y & DA+DC & F05 & 14.3 & 21.2 &  & S14 & 0.57 & 2014-06-06 \\ 
  WD2136+229 & 21:38:46.20 & +23:09:21.5 & N & DA & M99 & 15.2 & 38.0 &  & G11 & 0.10 & 2014-06-06 \\ 
  WD2149+021$^\dagger$ & 21:52:25.38 & +02:23:19.5 & N & DA & M99 & 13.2 & 25.1 & 2.8 & H08 & 0.06 & 2013-10-24 \\ 
  WD2213+317 & 22:15:06.96 & +31:58:40.2 & Y & DA+dM5 & S05 & 17.0 &  &  &  & 0.08 & 2014-06-06 \\ 
  WD2236+313 & 22:38:22.74 & +31:34:18.3 & N & DA & M99 & 14.5 &  &  &  & 0.09 & 2014-06-06 \\ 
  WD2306+124 & 23:08:35.00 & +12:45:40.2 & N & DA & M99 & 15.3 & 63.0 &  & G11 & 0.08 & 2014-06-06 \\ 
   \hline
\end{tabular}
}
\end{table*}

\begin{knitrout}
\definecolor{shadecolor}{rgb}{0.969, 0.969, 0.969}\color{fgcolor}\begin{figure}
\includegraphics[width=\maxwidth]{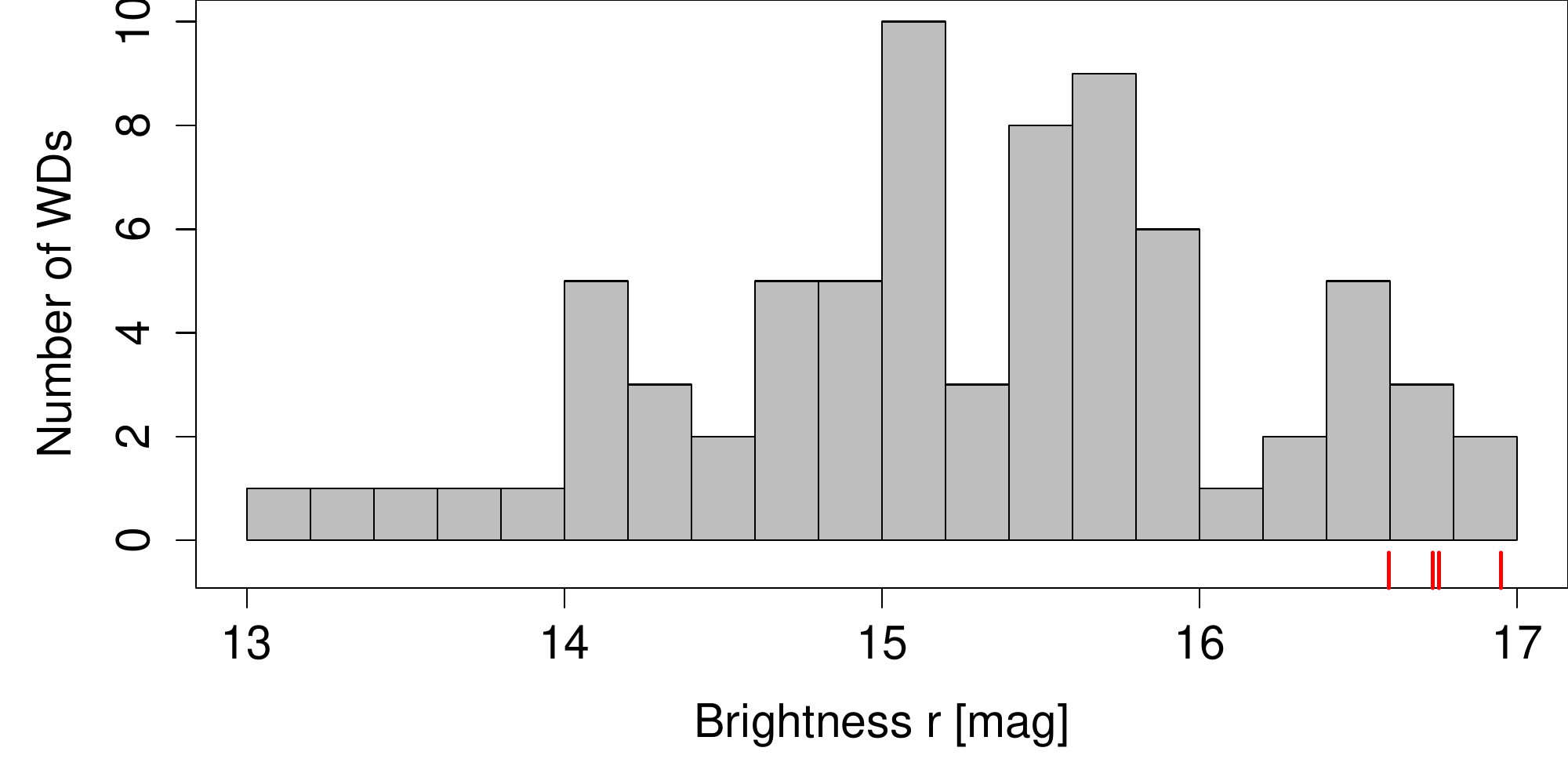} \caption[SDSS \textit{r} brightness  distribution of our WDs sample]{SDSS \textit{r} brightness  distribution of our WDs sample. The group consists
 of 74 WDs in total. Isolated DC type WDs are indicated with short red lines below the histogram.}\label{fig:histWD}
\end{figure}

\end{knitrout}

\begin{knitrout}
\definecolor{shadecolor}{rgb}{0.969, 0.969, 0.969}\color{fgcolor}\begin{figure*}
\includegraphics[width=\maxwidth]{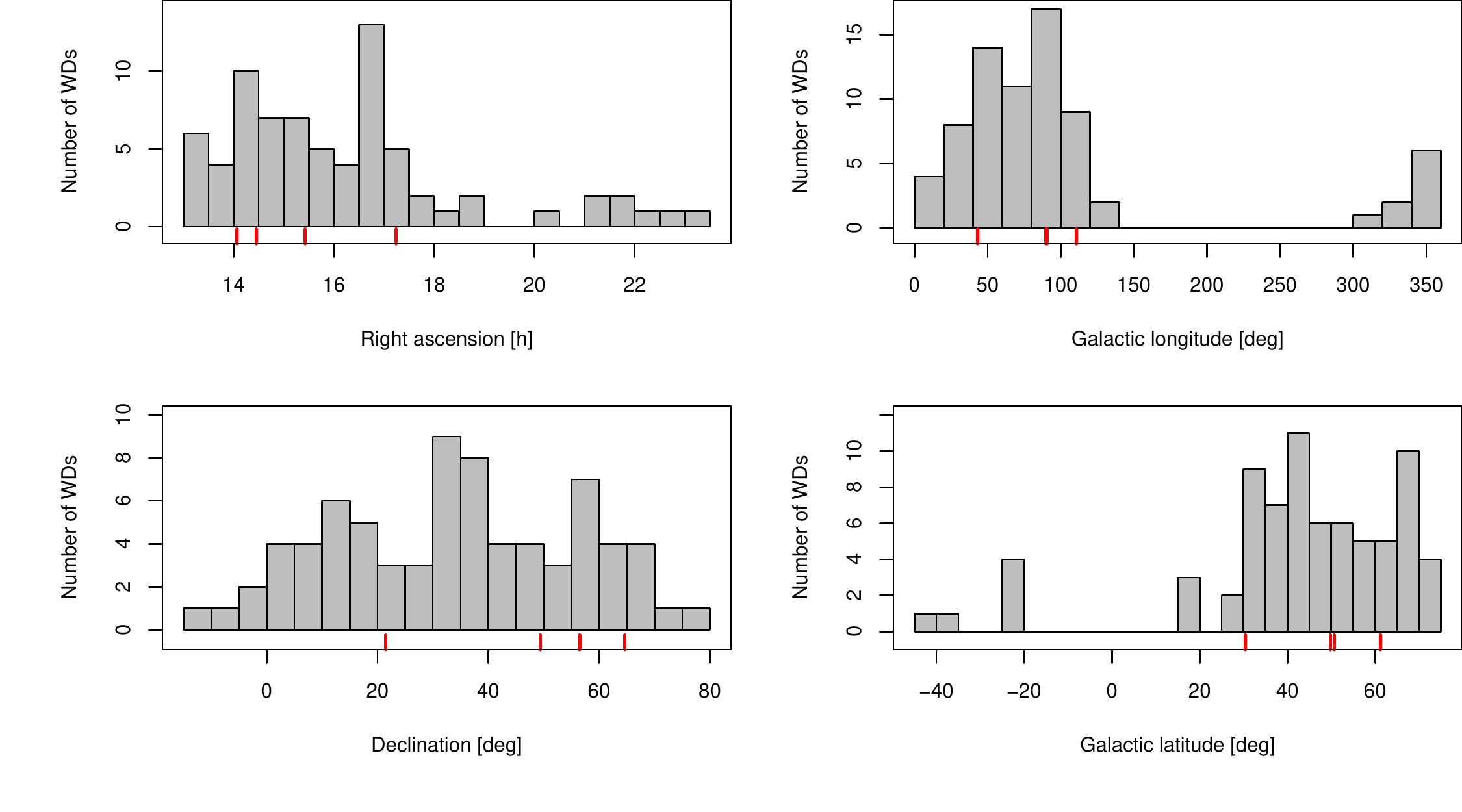} \caption[Distribution of WDs in the right ascension and the declination, as well as in
 the Galactic longitude and the Galactic latitude in the left and right columns,
 respectively]{Distribution of WDs in the right ascension and the declination, as well as in
 the Galactic longitude and the Galactic latitude in the left and right columns,
 respectively. There is a clear lack of WDs in our sample located in the Galactic
 latitude range between -20$\dg$ and +20$\dg$, i.e. in the location of the Galactic 
 disk.
 It is caused by the fact that surveys very often avoid this sky region. Isolated DC type 
 WDs are indicated with short red lines below each histogram.}\label{fig:histWD2}
\end{figure*}

\end{knitrout}

\section{Data analysis}
\label{sec:data}

The data were analysed with the standard RoboPol pipeline \citep{King2014}, whose output gives the normalised Stokes parameters $q=Q/I$ and $u=U/I$.
Because most of our objects have very low polarization degree, it is important to account for any instrumental polarization. To this end, we obtained 
22 measurements of the five zero-polarized standard stars given in Tab.~\ref{tab:polstan}. These measurements were performed during the time
span of the WDs observations. The resulting values of $q$ and $u$ with their 
corresponding errors are shown in Fig.~\ref{fig:zpolstan}. Almost all of the points locate in the second quarter, indicating some residual instrumental polarization. To compute the instrumental polarization, we calculated the mean $q$ and $u$ values from 22 measurements of $q$ and $u$ of the zero-standard stars and obtained $\overline{u}_{inst}=\ensuremath{-0.001}$ and $\overline{q}_{inst}=0.0033$, and
thus the instrumental PD given by the equation 
\begin{equation}
\label{eq:pd}
\rm{PD} = \sqrt{q^2 + u^2}
\end{equation}
equals to $0.34\% \pm 0.11\%$,
where the error is the standard deviation (Tab.~\ref{tab:polstan2}). The red cross in Fig.~\ref{fig:zpolstan} marks the position of the averaged $u$ and $q$ parameters, while the lengths of its arms correspond to the respective standard deviation. The average values of $\overline{u}_{inst}$
and $\overline{q}_{inst}$ were then subtracted from the WDs measured $u$ and $q$ values.
The results are given in Tab.~\ref{tab:results} and are shown in Fig.~\ref{fig:q_vs_u}. The standard deviations of $\overline{u}_{inst}$
and $\overline{q}_{inst}$ were propagated into the errors of the corrected 
WDs $q$ and $u$ values (corresponding $\sigma_\mathrm{q}$ and $\sigma_\mathrm{u}$ in
Tab.~\ref{tab:polstan2}). These values are also given in Tab.~\ref{tab:results}. The data corrected for instrumental polarization were later used to obtain the PD
according to Eq.~\ref{eq:pd}, while the corresponding error ($\sigma_{\rm{PD}}$) is calculated from  Eq. 5 in \cite{King2014}.
Almost all the observed WDs have PD lower than 1\%. 

For low polarization signal-to-noise ratio (PD / $\sigma_\mathrm{PD} < 3$)
distribution of the PD is not normal (Gaussian). Additionally,  
the values of PD must always be positive so their uncertainties
are not symmetric. This introduces a bias into any estimate. To deal with this
problem we applied the debias method described by \cite{Vaillancourt2006}.
First we checked the ratio between the PD and its error. In the case of PD/$\sigma_\mathrm{PD}$
lower than $\sqrt{2}$ the measured PD becomes 0.0 and only the upper sigma
PD is given (most of our sources). In case of $\sqrt{2}~<$~PD/$\sigma_\mathrm{PD}$~$~<1.7$ the PD value and upper error are 
unchanged, while the lower error was changed so the lower limit of PD is equal to zero. We have three such cases in our 
sample, i.e. WD1440-025, WD1553+353 and WD1738+669.
In case of PD/$\sigma_\mathrm{PD}$~$> 1.7$ the PD value is unchanged as well as the errors. Finally, in case of PD/$\sigma_\mathrm{PD}$~$>~3.0$ the 
PD value should be corrected according to the Eq.~12 of \cite{Vaillancourt2006}. However, we do not have any sources that 
fulfil the requirements of the last two mentioned cases. Obtained results
as well as the corrected PD$_\mathrm{c}$, $\sigma_\mathrm{PD_\mathrm{c}}^+$, $\sigma_\mathrm{PD_\mathrm{c}}^-$ are gathered in Tab.~\ref{tab:results}
in Sec.~\ref{sec:results}.

\begin{table}
\centering
\caption{Zero-polarized standard stars names, equatorial coordinates,
as well as their brightness in USNO-A2.0 R filter and spectral types (ST). BD~+28~4211
was classified as WD by \protect\cite{McCook1999}, according to \protect\cite{Gianninas2011}
its type is sdO.}
\begin{tabular}{lccrc}
\hline
\label{tab:polstan}
Name       & RA          & Dec               & R     &  ST \\
\hline
HD~94851    & 10:56:44.25 & --20:39:52.63 & 10.3   & A5V \\
HD~154892   & 17:07:41.31 & +15:12:37.61  & 7.9    & F8V \\
BD~+32~3739 & 20:12:02.15 & +32:47:43.71  & 9.0   & A0 \\
BD~+28~4211 & 21:51:11.02 & +28:51:50.35  & 10.9  & sdO \\
HD~212311   & 22:21:58.59 & +56:31:52.75  & 8.5   & A0 \\
\hline
\end{tabular}
\end{table}

\begin{knitrout}
\definecolor{shadecolor}{rgb}{0.969, 0.969, 0.969}\color{fgcolor}\begin{figure}
\includegraphics[width=\maxwidth]{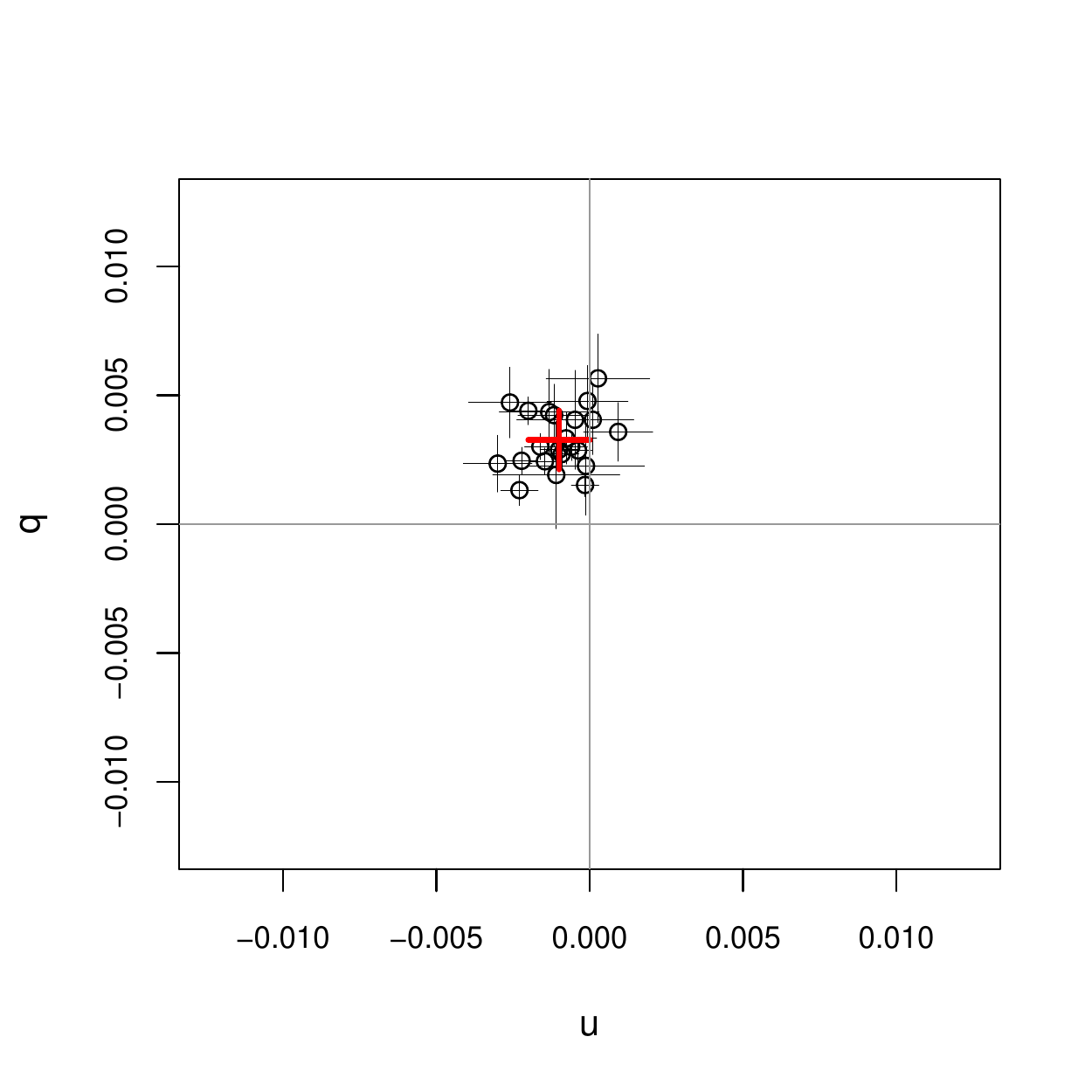} \caption[Diagram of ]{Diagram of $q$, $u$ parameters of observed zero-polarized standard stars. 
 The centre of the red cross corresponds to the mean values of q and u, while the lengths 
 of its arms correspond to the respective standard deviation.}\label{fig:zpolstan}
\end{figure}

\end{knitrout}

\begin{table}
 \centering
  \caption{Zero-polarized standard stars names, number of observations (NoO),
  the measured mean $u$ and mean $q$ values, together with the mean $u$ and $q$ values from all measurements. The PD calculated from the respective mean $u$ and mean $q$ is also given.}
  \begin{tabular}{lrccl}
  \hline
  \label{tab:polstan2}
Name & NoO & $\overline{u}$ & $\overline{q}$ & PD [\%]\\
\hline
HD~94851    & 2    
& \ensuremath{-3\times 10^{-4}}    
& 0.0022
& 0.22\\

HD~154892   & 2   
& \ensuremath{-0.0021}   
& 0.0029
& 0.36\\

BD~+32~3739 & 5 
& \ensuremath{-0.0011} &
0.0028
& 0.3\\

BD~+28~4211 & 12 
& \ensuremath{-8\times 10^{-4}} &
0.0038
& 0.39\\

HD~212311   & 1   
& \ensuremath{-0.0022}   
& 0.0025
& 0.33\\
\hline
\multicolumn{2}{l}{Mean value}            
& \ensuremath{-0.001}          
& 0.0033
& 0.34\\
\hline

   \end{tabular}
\end{table}

\section{Results}
\label{sec:results}

\subsection{Measured and corrected PD values}
The results of our analysis, i.e. the data set with $q$ and $u$ Stokes parameters, measured PD as well as PD corrected for low ratio of PD / $\sigma_\mathrm{PD}$ values along with their corresponding errors are presented in Tab.~\ref{tab:results}.

Our results do not contain the $q$ and $u$ values of two
well established zero-polarized standard white dwarfs, i.e. G~191-B2B
and GD319. It is caused by the fact that 
G~191-B2B was not visible at that time from Crete,
while GD319 was observed only once, but GD319 is a spectroscopic binary with the companion separation lower than 2-3 arcsec,
therefore proper aperture photometry of this object is very difficult with the instrument such as RoboPol.
However, in our other work on the calibration of the RINGO3 polarimeter,
we showed that the G~191-B2B is a very stable and good zero-polarized 
standard \citep{Slowikowska2016}.

\begin{knitrout}
\definecolor{shadecolor}{rgb}{0.969, 0.969, 0.969}\color{fgcolor}\begin{figure}
\includegraphics[width=\maxwidth]{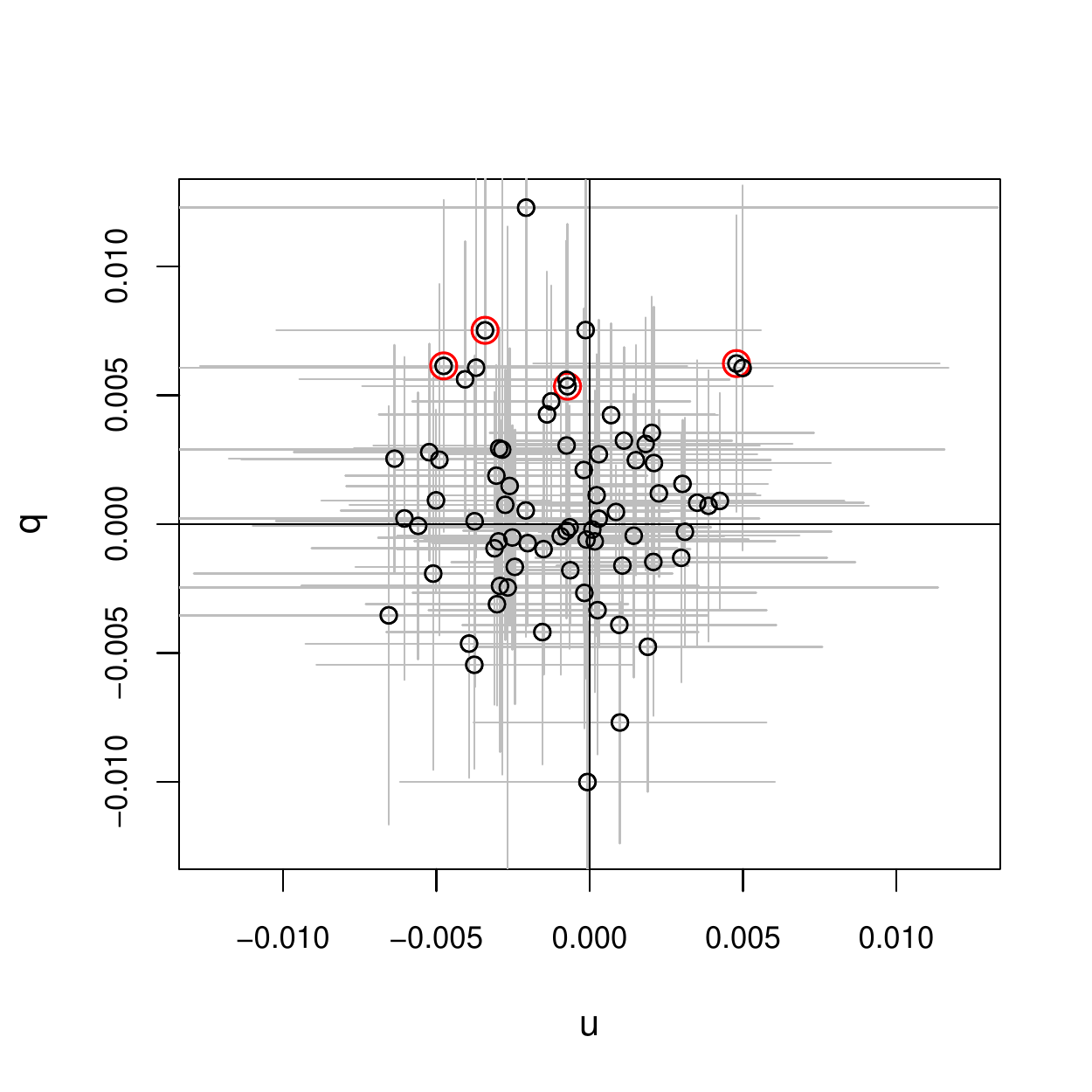} \caption{Diagram of $q$ and $u$ parameters of 74 observed WDs after correction for instrumental polarization. Red circles indicate Isolated WDs of DC type. The ranges of the figure axes are the same as in Fig.~\ref{fig:zpolstan} for easier comparison.}\label{fig:q_vs_u}
\end{figure}

\end{knitrout}

\begin{table*}
\centering
\caption{Results of polarimetric observations of WDs with the RoboPol polarimeter:
 WD name, MJD of observation, normalised Stokes parameters $q$ and $u$ with 
 instrumental polarization subtracted along with their respective errors, measured PD, $\sigma_\mathrm{PD}$, 
 corrected PD and and its upper and lower errors (see text for details). WDs common with  \citet{Schmidt1995} and \citet{Putney1997} 
 are marked with $^\dagger$.} 
\label{tab:results}
\scalebox{0.8}{
\begin{tabular}{lccccccccccc}
  \hline
WD Name & MJD & q & $\sigma_\mathrm{q}$ & u & $\sigma_\mathrm{u}$ & PD & $\sigma_\mathrm{PD}$ & PD / $\sigma_\mathrm{PD}$ & PD$_\mathrm{c}$ & $\sigma_\mathrm{PD_\mathrm{c}}^+$ & $\sigma_\mathrm{PD_\mathrm{c}}^-$ \\ 
  \hline
WD1257+037$^\dagger$ & 56799.82 & 0.0032 & 0.0036 & 0.0011 & 0.0035 & 0.34 & 0.36 & 0.95 & 0.00 & 0.36 & 0.00 \\ 
  WD1258+593 & 56799.99 & 0.0009 & 0.0035 & $-$0.0050 & 0.0037 & 0.51 & 0.37 & 1.37 & 0.00 & 0.37 & 0.00 \\ 
  WD1259+674 & 56799.87 & $-$0.0005 & 0.0044 & $-$0.0025 & 0.0044 & 0.26 & 0.44 & 0.59 & 0.00 & 0.44 & 0.00 \\ 
  WD1310+583$^\dagger$ & 56813.79 & 0.0019 & 0.0043 & $-$0.0030 & 0.0049 & 0.36 & 0.48 & 0.75 & 0.00 & 0.48 & 0.00 \\ 
  WD1317+453$^\dagger$ & 56808.77 & 0.0048 & 0.0045 & $-$0.0013 & 0.0045 & 0.49 & 0.45 & 1.09 & 0.00 & 0.45 & 0.00 \\ 
  WD1319+466$^\dagger$ & 56813.80 & 0.0028 & 0.0042 & $-$0.0052 & 0.0044 & 0.59 & 0.44 & 1.36 & 0.00 & 0.44 & 0.00 \\ 
  WD1334+070 & 56813.81 & 0.0024 & 0.0061 & 0.0021 & 0.0058 & 0.32 & 0.59 & 0.53 & 0.00 & 0.59 & 0.00 \\ 
  WD1344+106 & 56813.81 & 0.0025 & 0.0068 & $-$0.0049 & 0.0065 & 0.55 & 0.65 & 0.84 & 0.00 & 0.65 & 0.00 \\ 
  WD1344+572$^\dagger$ & 56808.78 & $-$0.0031 & 0.0039 & $-$0.0030 & 0.0043 & 0.43 & 0.41 & 1.06 & 0.00 & 0.41 & 0.00 \\ 
  WD1349+144 & 56813.82 & $-$0.0039 & 0.0053 & 0.0010 & 0.0051 & 0.40 & 0.52 & 0.77 & 0.00 & 0.52 & 0.00 \\ 
  WD1401+005 & 56813.83 & $-$0.0005 & 0.0054 & $-$0.0009 & 0.0053 & 0.11 & 0.53 & 0.20 & 0.00 & 0.53 & 0.00 \\ 
  WD1402+649 & 56808.80 & 0.0061 & 0.0064 & $-$0.0048 & 0.0079 & 0.78 & 0.70 & 1.10 & 0.00 & 0.70 & 0.00 \\ 
  WD1407+374 & 56813.83 & 0.0029 & 0.0047 & $-$0.0030 & 0.0047 & 0.42 & 0.47 & 0.88 & 0.00 & 0.47 & 0.00 \\ 
  WD1407+425$^\dagger$ & 56808.96 & 0.0001 & 0.0064 & $-$0.0038 & 0.0065 & 0.38 & 0.65 & 0.58 & 0.00 & 0.65 & 0.00 \\ 
  WD1408+323$^\dagger$ & 56813.84 & $-$0.0002 & 0.0031 & 0.0001 & 0.0030 & 0.02 & 0.31 & 0.07 & 0.00 & 0.31 & 0.00 \\ 
  WD1415+132 & 56813.84 & 0.0008 & 0.0055 & 0.0035 & 0.0054 & 0.36 & 0.54 & 0.66 & 0.00 & 0.54 & 0.00 \\ 
  WD1420+518 & 56813.85 & $-$0.0027 & 0.0053 & $-$0.0002 & 0.0056 & 0.27 & 0.53 & 0.51 & 0.00 & 0.53 & 0.00 \\ 
  WD1421+318 & 56813.86 & $-$0.0042 & 0.0051 & $-$0.0015 & 0.0051 & 0.45 & 0.51 & 0.87 & 0.00 & 0.51 & 0.00 \\ 
  WD1422+497 & 56813.87 & $-$0.0046 & 0.0052 & $-$0.0039 & 0.0053 & 0.61 & 0.53 & 1.16 & 0.00 & 0.53 & 0.00 \\ 
  WD1425+495 & 56808.81 & 0.0054 & 0.0063 & $-$0.0007 & 0.0067 & 0.54 & 0.63 & 0.86 & 0.00 & 0.63 & 0.00 \\ 
  WD1428+373 & 56813.88 & 0.0035 & 0.0053 & 0.0020 & 0.0053 & 0.41 & 0.53 & 0.77 & 0.00 & 0.53 & 0.00 \\ 
  WD1429+373 & 56808.98 & $-$0.0048 & 0.0056 & 0.0019 & 0.0057 & 0.51 & 0.56 & 0.91 & 0.00 & 0.56 & 0.00 \\ 
  WD1434+328 & 56813.89 & 0.0015 & 0.0053 & $-$0.0026 & 0.0053 & 0.30 & 0.53 & 0.56 & 0.00 & 0.53 & 0.00 \\ 
  WD1440$-$025 & 56813.90 & $-$0.0100 & 0.0062 & $-$0.0001 & 0.0061 & 1.00 & 0.62 & 1.60 & 1.00 & 0.62 & 1.00 \\ 
  WD1446+286$^\dagger$ & 56813.90 & $-$0.0010 & 0.0049 & $-$0.0015 & 0.0048 & 0.18 & 0.48 & 0.37 & 0.00 & 0.48 & 0.00 \\ 
  WD1447+049 & 56813.92 & 0.0025 & 0.0045 & 0.0015 & 0.0044 & 0.29 & 0.44 & 0.65 & 0.00 & 0.44 & 0.00 \\ 
  WD1449+168 & 56813.93 & $-$0.0017 & 0.0053 & $-$0.0024 & 0.0052 & 0.30 & 0.52 & 0.56 & 0.00 & 0.52 & 0.00 \\ 
  WD1503$-$093 & 56813.93 & 0.0007 & 0.0053 & 0.0039 & 0.0052 & 0.39 & 0.52 & 0.75 & 0.00 & 0.52 & 0.00 \\ 
  WD1507$-$105 & 56813.94 & 0.0011 & 0.0055 & 0.0002 & 0.0053 & 0.11 & 0.55 & 0.21 & 0.00 & 0.55 & 0.00 \\ 
  WD1515+668 & 56808.99 & 0.0005 & 0.0049 & $-$0.0021 & 0.0060 & 0.22 & 0.60 & 0.36 & 0.00 & 0.60 & 0.00 \\ 
  WD1518$-$003 & 56808.96 & 0.0043 & 0.0055 & $-$0.0014 & 0.0055 & 0.45 & 0.55 & 0.81 & 0.00 & 0.55 & 0.00 \\ 
  WD1524+566 & 56808.83 & 0.0062 & 0.0058 & 0.0048 & 0.0066 & 0.79 & 0.61 & 1.29 & 0.00 & 0.61 & 0.00 \\ 
  WD1525+257 & 56813.96 & $-$0.0009 & 0.0061 & $-$0.0031 & 0.0060 & 0.32 & 0.60 & 0.54 & 0.00 & 0.60 & 0.00 \\ 
  WD1527+090 & 56808.89 & 0.0042 & 0.0036 & 0.0007 & 0.0035 & 0.43 & 0.36 & 1.21 & 0.00 & 0.36 & 0.00 \\ 
  WD1534+503 & 56809.00 & $-$0.0019 & 0.0076 & $-$0.0051 & 0.0078 & 0.55 & 0.78 & 0.70 & 0.00 & 0.78 & 0.00 \\ 
  WD1538+269 & 56798.00 & $-$0.0007 & 0.0032 & $-$0.0020 & 0.0031 & 0.22 & 0.31 & 0.69 & 0.00 & 0.31 & 0.00 \\ 
  WD1538+333 & 56809.01 & $-$0.0024 & 0.0064 & $-$0.0029 & 0.0065 & 0.38 & 0.65 & 0.58 & 0.00 & 0.65 & 0.00 \\ 
  WD1548+149 & 56809.01 & 0.0056 & 0.0054 & $-$0.0041 & 0.0054 & 0.69 & 0.54 & 1.29 & 0.00 & 0.54 & 0.00 \\ 
  WD1553+353$^\dagger$ & 56808.98 & $-$0.0077 & 0.0047 & 0.0010 & 0.0048 & 0.78 & 0.47 & 1.65 & 0.78 & 0.47 & 0.78 \\ 
  WD1601+581$^\dagger$ & 56808.84 & 0.0016 & 0.0025 & 0.0030 & 0.0028 & 0.34 & 0.27 & 1.25 & 0.00 & 0.27 & 0.00 \\ 
  WD1606+422$^\dagger$ & 56808.90 & $-$0.0003 & 0.0034 & $-$0.0008 & 0.0034 & 0.08 & 0.34 & 0.24 & 0.00 & 0.34 & 0.00 \\ 
  WD1610+166 & 56809.02 & $-$0.0025 & 0.0140 & $-$0.0027 & 0.0140 & 0.36 & 1.40 & 0.26 & 0.00 & 1.40 & 0.00 \\ 
  WD1614+160 & 56813.99 & 0.0075 & 0.0058 & $-$0.0001 & 0.0057 & 0.75 & 0.58 & 1.29 & 0.00 & 0.58 & 0.00 \\ 
  WD1630+089 & 56808.97 & 0.0008 & 0.0052 & $-$0.0028 & 0.0051 & 0.29 & 0.51 & 0.56 & 0.00 & 0.51 & 0.00 \\ 
  WD1631+396 & 56808.90 & $-$0.0006 & 0.0054 & $-$0.0001 & 0.0053 & 0.06 & 0.54 & 0.11 & 0.00 & 0.54 & 0.00 \\ 
  WD1632+177$^\dagger$ & 56808.89 & 0.0005 & 0.0018 & 0.0009 & 0.0017 & 0.10 & 0.18 & 0.56 & 0.00 & 0.18 & 0.00 \\ 
  WD1636+160 & 56813.99 & 0.0031 & 0.0064 & $-$0.0008 & 0.0063 & 0.31 & 0.64 & 0.50 & 0.00 & 0.64 & 0.00 \\ 
  WD1636+351 & 56808.92 & $-$0.0004 & 0.0055 & 0.0014 & 0.0054 & 0.15 & 0.54 & 0.28 & 0.00 & 0.54 & 0.00 \\ 
  WD1637+335$^\dagger$ & 56808.91 & $-$0.0001 & 0.0047 & $-$0.0007 & 0.0046 & 0.07 & 0.46 & 0.14 & 0.00 & 0.46 & 0.00 \\ 
  WD1639+537$^\dagger$ & 56808.85 & $-$0.0003 & 0.0044 & 0.0031 & 0.0048 & 0.31 & 0.48 & 0.65 & 0.00 & 0.48 & 0.00 \\ 
  WD1641+387$^\dagger$ & 56808.92 & 0.0031 & 0.0049 & 0.0018 & 0.0048 & 0.36 & 0.49 & 0.74 & 0.00 & 0.49 & 0.00 \\ 
  WD1643+143 & 56808.93 & $-$0.0033 & 0.0056 & 0.0003 & 0.0055 & 0.33 & 0.56 & 0.60 & 0.00 & 0.56 & 0.00 \\ 
  WD1647+375$^\dagger$ & 56808.95 & $-$0.0001 & 0.0052 & $-$0.0056 & 0.0054 & 0.56 & 0.54 & 1.04 & 0.00 & 0.54 & 0.00 \\ 
  WD1654+637 & 56814.01 & 0.0002 & 0.0063 & $-$0.0060 & 0.0076 & 0.60 & 0.76 & 0.80 & 0.00 & 0.76 & 0.00 \\ 
  WD1655+215$^\dagger$ & 56808.86 & 0.0012 & 0.0032 & 0.0023 & 0.0032 & 0.26 & 0.32 & 0.80 & 0.00 & 0.32 & 0.00 \\ 
  WD1658+440$^\dagger$ & 56808.94 & 0.0002 & 0.0050 & 0.0003 & 0.0052 & 0.04 & 0.52 & 0.07 & 0.00 & 0.52 & 0.00 \\ 
  WD1659+303$^\dagger$ & 56808.94 & 0.0027 & 0.0052 & 0.0003 & 0.0052 & 0.27 & 0.52 & 0.52 & 0.00 & 0.52 & 0.00 \\ 
  WD1706+332 & 56808.91 & 0.0061 & 0.0071 & 0.0050 & 0.0067 & 0.78 & 0.69 & 1.13 & 0.00 & 0.69 & 0.00 \\ 
  WD1712+215 & 56808.88 & 0.0075 & 0.0071 & $-$0.0034 & 0.0068 & 0.83 & 0.71 & 1.17 & 0.00 & 0.71 & 0.00 \\ 
  WD1713+695$^\dagger$ & 56808.95 & $-$0.0007 & 0.0022 & $-$0.0030 & 0.0025 & 0.30 & 0.25 & 1.23 & 0.00 & 0.25 & 0.00 \\ 
  WD1723+563 & 56809.04 & $-$0.0015 & 0.0060 & 0.0021 & 0.0066 & 0.25 & 0.64 & 0.40 & 0.00 & 0.64 & 0.00 \\ 
  WD1734+575 & 56814.02 & 0.0029 & 0.0126 & $-$0.0029 & 0.0144 & 0.41 & 1.35 & 0.30 & 0.00 & 1.35 & 0.00 \\ 
  WD1738+669 & 56814.03 & $-$0.0055 & 0.0040 & $-$0.0038 & 0.0051 & 0.66 & 0.44 & 1.50 & 0.66 & 0.44 & 0.66 \\ 
  WD1827+778 & 56814.04 & 0.0061 & 0.0074 & $-$0.0037 & 0.0113 & 0.71 & 0.86 & 0.83 & 0.00 & 0.86 & 0.00 \\ 
  WD1833+644 & 56814.05 & 0.0025 & 0.0044 & $-$0.0064 & 0.0054 & 0.69 & 0.53 & 1.30 & 0.00 & 0.53 & 0.00 \\ 
  WD1842+412 & 56809.07 & $-$0.0007 & 0.0059 & 0.0002 & 0.0059 & 0.07 & 0.59 & 0.12 & 0.00 & 0.59 & 0.00 \\ 
  WD2006+615 & 56809.07 & $-$0.0035 & 0.0081 & $-$0.0066 & 0.0104 & 0.74 & 0.99 & 0.75 & 0.00 & 0.99 & 0.00 \\ 
  WD2058+083 & 56809.08 & 0.0056 & 0.0054 & $-$0.0008 & 0.0053 & 0.57 & 0.54 & 1.05 & 0.00 & 0.54 & 0.00 \\ 
  WD2126+734$^\dagger$ & 56814.06 & $-$0.0016 & 0.0019 & 0.0011 & 0.0021 & 0.19 & 0.20 & 0.98 & 0.00 & 0.20 & 0.00 \\ 
  WD2136+229 & 56814.07 & 0.0009 & 0.0042 & 0.0042 & 0.0041 & 0.43 & 0.41 & 1.07 & 0.00 & 0.41 & 0.00 \\ 
  WD2149+021$^\dagger$ & 56589.88 & $-$0.0018 & 0.0021 & $-$0.0006 & 0.0021 & 0.19 & 0.21 & 0.90 & 0.00 & 0.21 & 0.00 \\ 
  WD2213+317 & 56814.08 & 0.0123 & 0.0159 & $-$0.0021 & 0.0154 & 1.25 & 1.59 & 0.78 & 0.00 & 1.59 & 0.00 \\ 
  WD2236+313 & 56814.09 & $-$0.0013 & 0.0048 & 0.0030 & 0.0048 & 0.33 & 0.48 & 0.68 & 0.00 & 0.48 & 0.00 \\ 
  WD2306+124 & 56814.09 & 0.0021 & 0.0063 & $-$0.0002 & 0.0061 & 0.21 & 0.63 & 0.34 & 0.00 & 0.63 & 0.00 \\ 
   \hline
\end{tabular}
}
\end{table*}

\begin{knitrout}
\definecolor{shadecolor}{rgb}{0.969, 0.969, 0.969}\color{fgcolor}\begin{figure}
\includegraphics[width=\maxwidth]{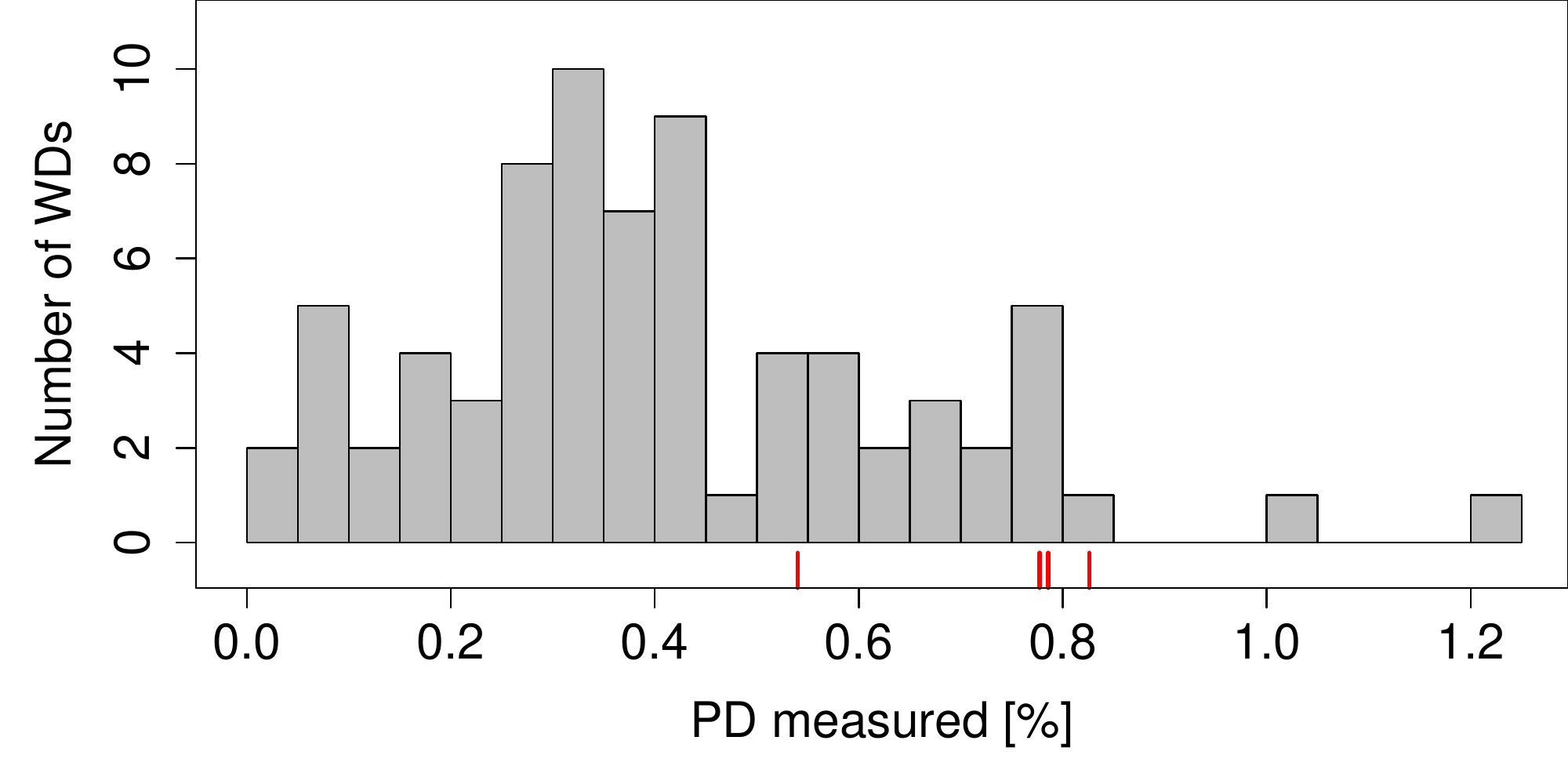} \caption[Distribution of measured PD regardless of the spectral type of WDs]{Distribution of measured PD regardless of the spectral type of WDs. Isolated DC type WDs are indicated with short red lines below the histogram.}\label{fig:pd_hist}
\end{figure}

\end{knitrout}

\begin{knitrout}
\definecolor{shadecolor}{rgb}{0.969, 0.969, 0.969}\color{fgcolor}\begin{figure}
\includegraphics[width=\maxwidth]{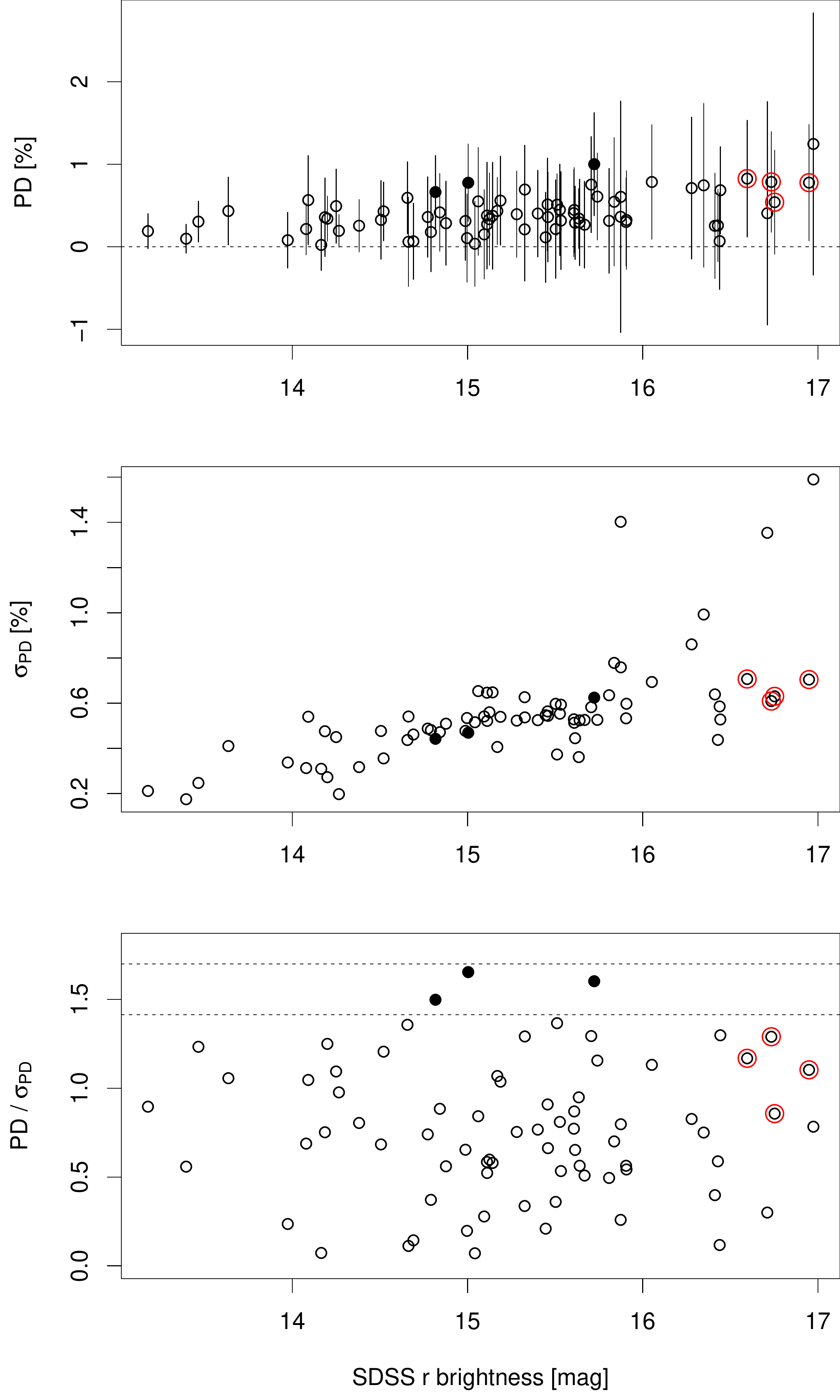} \caption{Calculated PD, $\sigma_\mathrm{PD}$ as well as the PD / $\sigma_\mathrm{PD}$ as a function
 of SDSS \textit{r} brightness in the top, middle, and bottom panel, respectively.
 Black points denote measurements with PD / $\sigma_\mathrm{PD}$ $> \sqrt{2}$, while red circles
 indicate isolated DC type WDs. The horizontal lines in the bottom panel show $\sqrt{2}$
 and 1.7 levels (see text for details as well as Tab.~\ref{tab:results}).}\label{fig:correlations}
\end{figure}

\end{knitrout}

\begin{knitrout}
\definecolor{shadecolor}{rgb}{0.969, 0.969, 0.969}\color{fgcolor}\begin{figure*}
\includegraphics[width=\maxwidth]{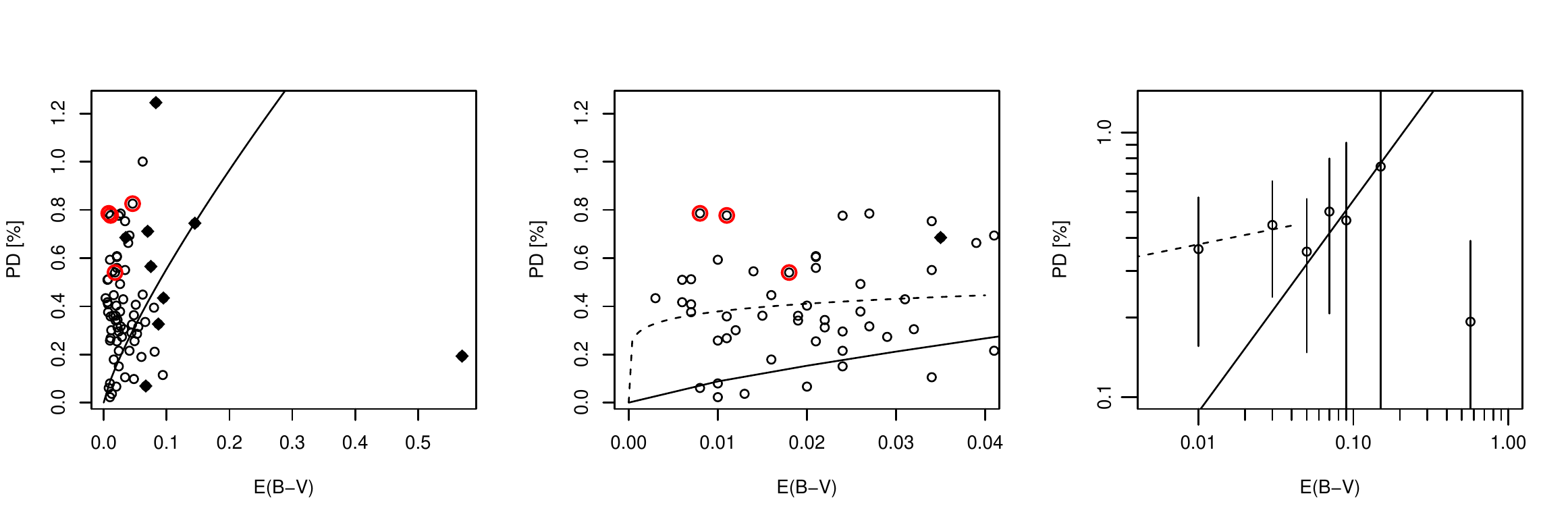} \caption{Dependence of the PD on the interstellar extinction for observed WDs.
 Values of E(B-V) were obtained from \citet{Schlafly2011} using the NASA/IPAC Extragalactic Database \citep{NED2009}. 
 The left and middle panels differ only with the x--axis range, while the right panel
 (log-log) shows data averaged in extinction bins of size 0.02. 
 The red circles indicate isolated DC type WDs, while the black rhombuses indicate 
 the WDs with the Galactic latitude between -30$\dg$ and +30$\dg$.
 The solid lines show expected dependence of the PD on E(B-V) given by 
 Eq.~(\ref{eq:ism}). The dashed lines show the dependence of the PD on E(B-V)
 for the extinction range between 0.0--0.04 with the Eq.~(\ref{eq:ism.low}). 
 Object WD2126+734 with the highest E(B-V) = 0.57 has PD equal to 0.19\%.
 The error bars are omitted for clarity in the left and middle panels. 
 In the right panel the errors are taken as standard deviations of measurements that fall into particular bin, apart for the last two bins. In those bins only single measurements were available and their individual $\sigma_\mathrm{PD}$ were taken as the errors.}\label{fig:pd_vs_ebv}
\end{figure*}

\end{knitrout}

\begin{knitrout}
\definecolor{shadecolor}{rgb}{0.969, 0.969, 0.969}\color{fgcolor}\begin{figure}
\includegraphics[width=\maxwidth]{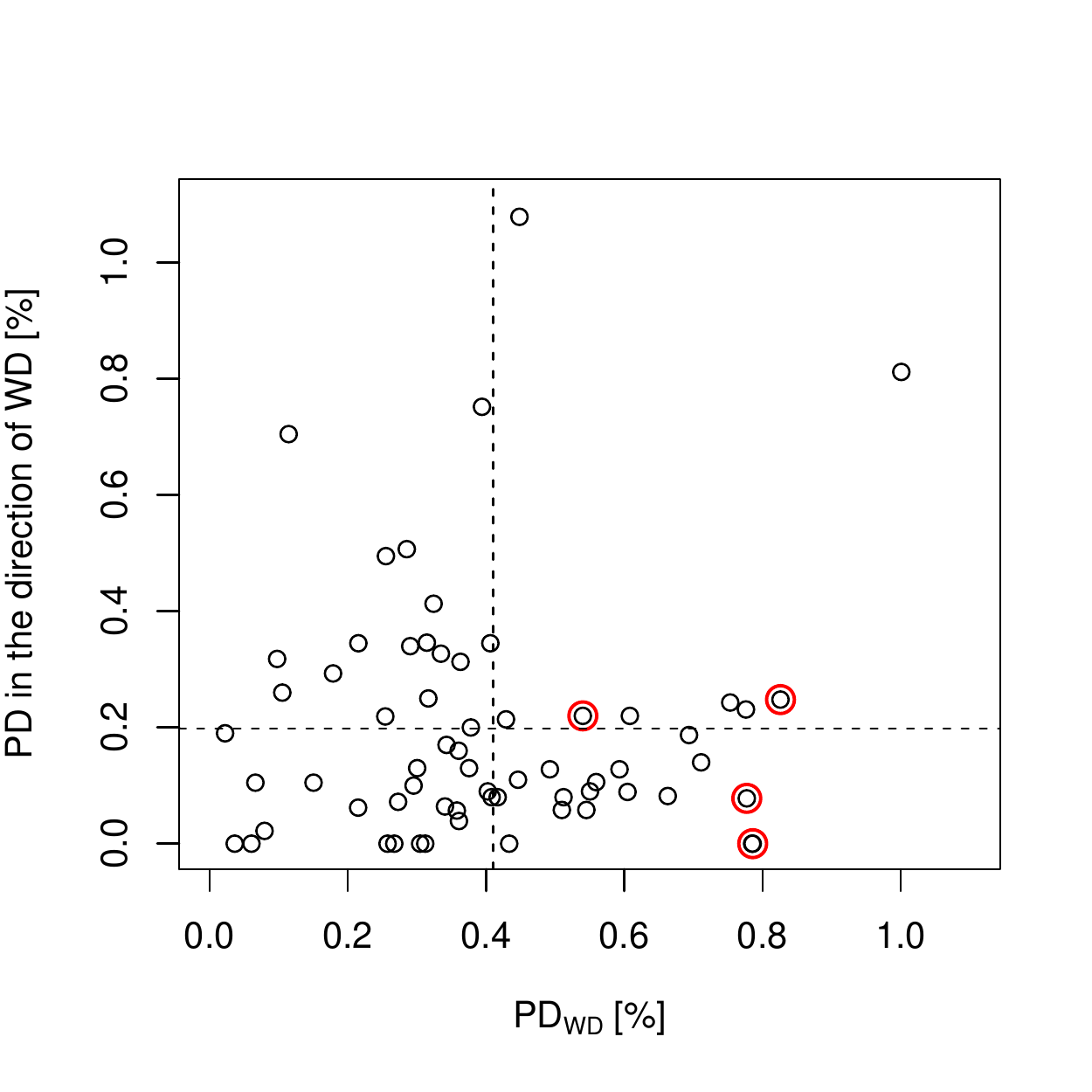} \caption[Measured PD values from \citet{Berdyugin2014} in the direction of a given WD within 
 the 3]{Measured PD values from \citet{Berdyugin2014} in the direction of a given WD within 
 the 3$\dg$ radius vs. WDs measured PD. Isolated DC type WDs are marked with red circles.  WD1440$-$025 that has PD$\sim$1\%, is classified as DA+dMe. Dashed lines denote
 respective mean PD values. The error bars are omitted for clarity.}\label{fig:PD_gal_vs_PD_WD}
\end{figure}

\end{knitrout}

\begin{knitrout}
\definecolor{shadecolor}{rgb}{0.969, 0.969, 0.969}\color{fgcolor}\begin{figure*}
\includegraphics[width=\maxwidth]{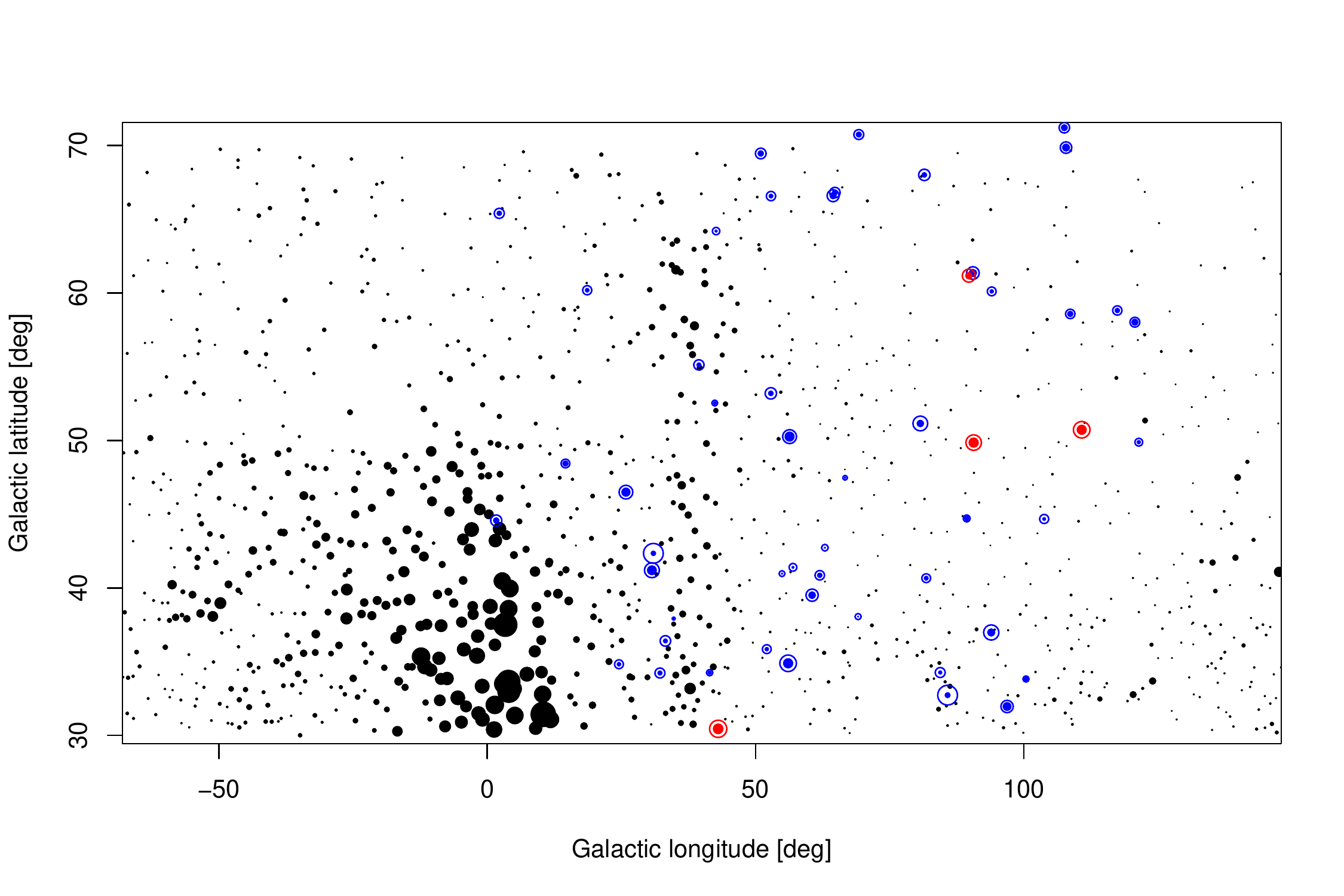} \caption{The spatial distribution
 of our WDs sample in galactic coordinates
 in comparison to \citet{Berdyugin2014} measurements. Size of the filled
 black circles corresponds to the PD value measured by \citet{Berdyugin2014},
 whereas the diameter of colour circles corresponds to the PD value measured 
 by us, red circles indicate the isolated DC WDs, whereas the blue circles indicate all
 other WDs. Open circles denote upper limit of measured PD, i.e. PD + $\sigma_\mathrm{PD}$. As can be seen, there are regions in
 the Galaxy that show higher polarization degree, however DC WDs from our
 sample are not located in these regions, therefore their PD most likely is not 
 influenced by the background polarization.}\label{fig:gal_pol}
\end{figure*}

\end{knitrout}

\begin{knitrout}
\definecolor{shadecolor}{rgb}{0.969, 0.969, 0.969}\color{fgcolor}\begin{figure}
\includegraphics[width=\maxwidth]{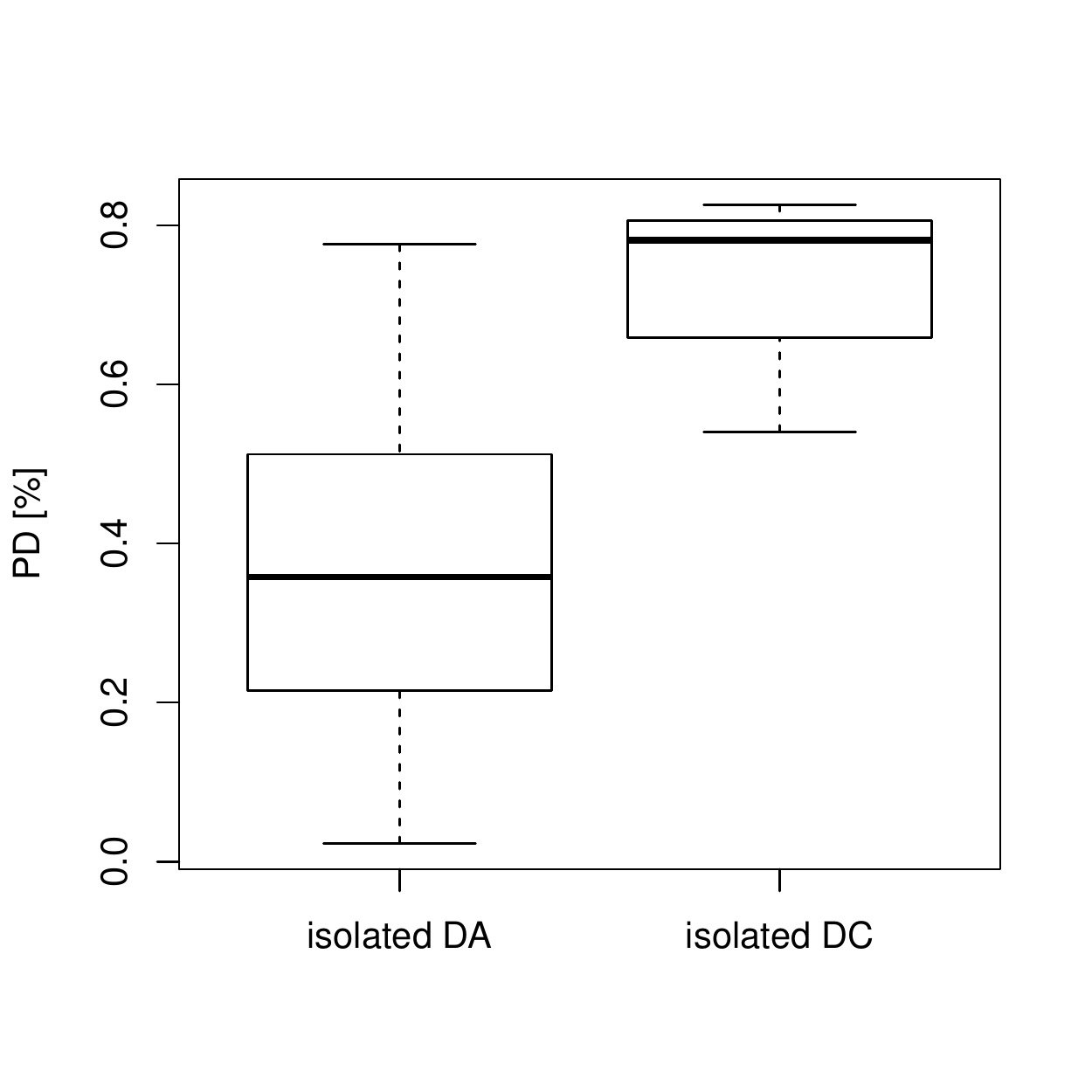} \caption[Comparison of PD of isolated DA and isolated 
 DC type white dwarfs in form of a box plot]{Comparison of PD of isolated DA and isolated 
 DC type white dwarfs in form of a box plot. The thick black line
 inside the rectangle denotes the median value of the sample.
 The rectangle covers 50\% 
 of the points lying between 1st and 3rd quartiles. The upper and the 
 lower whiskers denote either extremal
 value from the dataset (maximum or minimum, respectively) lying in 
 the 1.5 IQR (interquartile range) distance from
 the rectangle or 1.5 IQR itself, whichever is closer. These two groups are clearly different. The DC type WDs show 
 higher PD. One should be aware that
 there are only 4 isolated   
 white dwarfs of DC 
 spectral type in the sample.}\label{fig:da_vs_dc}
\end{figure}

\end{knitrout}

\begin{knitrout}
\definecolor{shadecolor}{rgb}{0.969, 0.969, 0.969}\color{fgcolor}\begin{figure}
\includegraphics[width=\maxwidth]{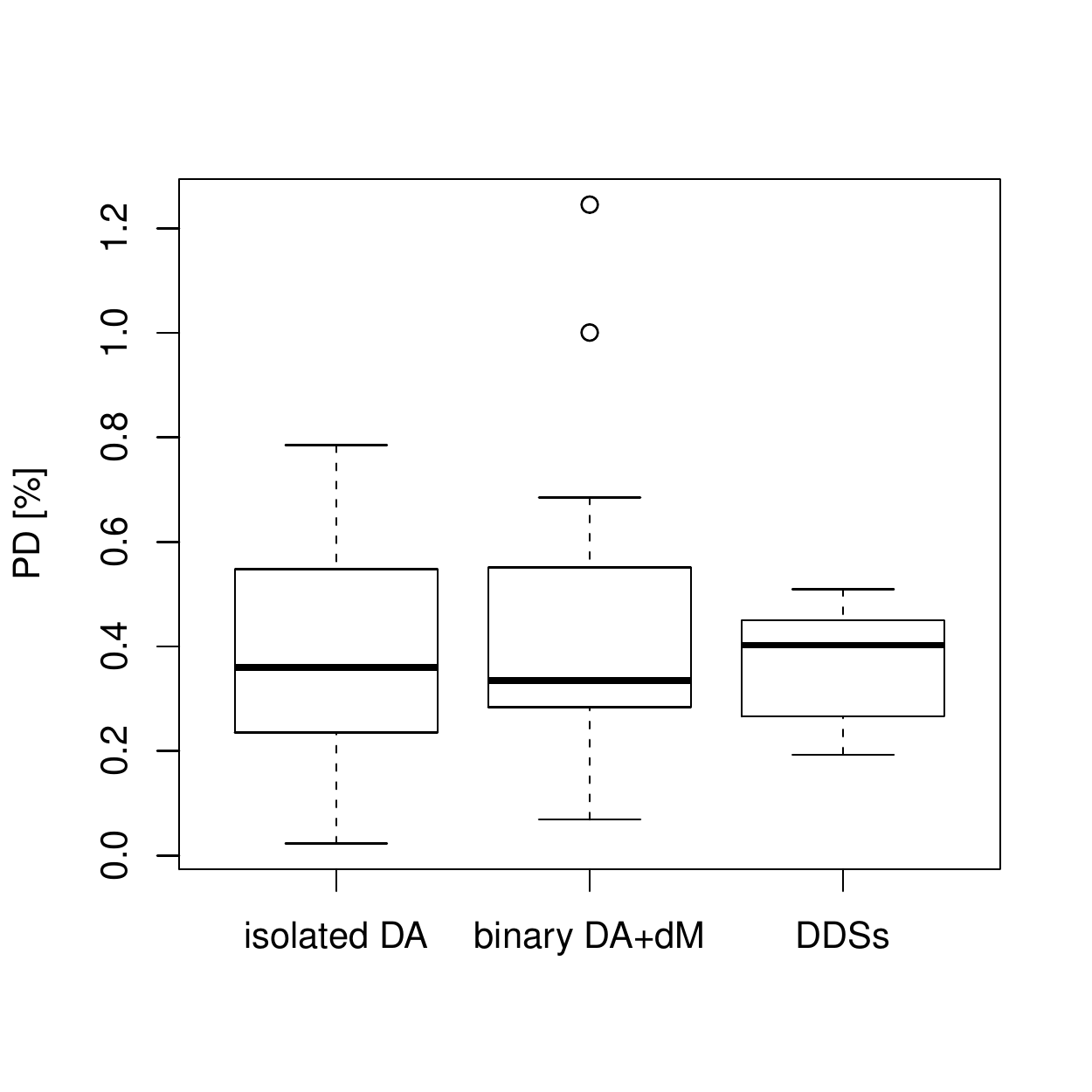} \caption{Comparison of PD of isolated DA type white dwarfs, DA+dM binaries and double degenerate systems (DDSs) 
 presented as a box plot similar as in Fig.~\ref{fig:da_vs_dc}. 
 Outliers, i.e. WD1440$-$025 and WD2213+317, lying outside the whiskers range are plotted individually.  
 }\label{fig:iso_vs_bin}
\end{figure}

\end{knitrout}

The distribution of measured PD  regardless of WD spectral type or binarity
is shown in Fig.~\ref{fig:pd_hist}. 
The plots of PD, $\sigma_\mathrm{PD}$ and PD / $\sigma_\mathrm{PD}$ with respect to SDSS
\textit{r} brightness are shown in Fig.~\ref{fig:correlations}.
A correlation analysis between these parameters yields 
the following correlation coefficients: 0.48 (p-value = $\ensuremath{1.665\times 10^{-5}}$),
0.66 (p-value = $\ensuremath{1.287\times 10^{-10}}$) and 0  
(p-value = $0.996$), respectively. This suggests that there might be a dependence on the brightness
in both cases of PD and $\sigma_\mathrm{PD}$.

\subsection{PD dependence on the interstellar extinction}

Additionally, we checked correlations between polarimetric results and location \textit{(l, b)} in the Galaxy, but we found none. However, the interstellar reddening E(B-V) depends on the location of the source in the Galaxy. Therefore, we also studied the PD--extinction relationship as described by \citet{Fosalba2002} who found it to be of the form of
\begin{equation}
\label{eq:ism}
\rm{PD_{max, ISM}}[\%] = 3.5~E(B-V)^{0.8}~.
\end{equation}
Fig.~\ref{fig:pd_vs_ebv} shows the PD as a function of E(B--V) for all WDs in our sample.
The relation described by the Eq.~(\ref{eq:ism}) is shown in the figure with the black solid lines. Our data range
from 0.003 to 0.57 in terms of reddening, because most of the stars are closer than 500 pc. They also have small
PD (<1\%). \citet{Fosalba2002} E(B-V) binned data range from 0.1 to 1.0, with one exception of E(B--V) = 0.05, as it is shown in their log--log scale Fig. 4. For comparison, we also binned our data in extinction with the bin size of 0.02. Resulting plot is shown in the right panel of Fig.~\ref{fig:pd_vs_ebv}, where we have 7 points that are averages of 26, 22, 12, 7, 5, 1, 1 measurements, respectively. Most of the points are located in the lower E(B--V)-bins. The sixth and seventh bins correspond to WD2006+615 and WD2126+734 with
E(B--V) values of 0.145 and 0.57 and PD of 0.74\% and 0.19\%, respectively. PD errors in the right panel of Fig.~\ref{fig:pd_vs_ebv}
are taken as standard deviations of measurements that fell into particular bin but for the last two bins individual $\sigma_\mathrm{PD}$ were taken. It can be seen that
the first two values (first and second bin) significantly diverge from model (Eq.~\ref{eq:ism}, shown as black solid line in all panels of Fig.~\ref{fig:pd_vs_ebv}). It is worth noticing that most of extinction values of our WDs, 48 out of 74 measurements, fall into the first two bins with E(B--V) below 0.04.  Therefore, we fitted the $a \cdot \mathrm{E(V-B)} ^ b$ model to that data and got the following relation in the  extinction range from 0.0 up to 0.04:
\begin{equation}
\label{eq:ism.low}
\rm{PD_{max, ISM}}[\%] = 0.65~E(B-V)^{0.12}
\end{equation}
This dependence is plotted as dashed line in the middle and right panels of Fig.~\ref{fig:pd_vs_ebv}. 

\subsection{Comparision of our measurments with the PD survey}
We also compared the measured PD of WDs in our sample with the catalogue values of PD obtained by \cite{Berdyugin2014}, who collected PD measurements 
of $\sim$3600 stars within the distance of 600~pc. To get galactic polarization
maps they took the data in the V band. They excluded variable
and peculiar stars which might posses intrinsic polarization. More than
60\% of stars in their sample show polarization below 0.5\%. Only around
200 stars have PD higher than 0.5\%, up to values exceeding 1\%.
We have the distance value for 51 WDs in our sample,
where the closest one is
13.2~pc away and the farthest has a distance of 
463~pc (Tab.~\ref{tab:log}).
For each WD in our sample we searched for neighbouring stars
in the VizieR Online Data Catalog "Polarization at high galactic 
latitude" \citep{Berdyugin2014} within the radius of 3$\dg$.
We found 64 cases that allowed us to compare our WD measurements with catalogue PD values
what is shown in Fig.~\ref{fig:PD_gal_vs_PD_WD}.  We can see that an average PD from
the \cite{Berdyugin2014} measurements in the direction to WDs from our sample
is on the level of 
$0.2\% \pm 0.03\%$, 
whereas the mean PD of these
64 WDs is 
$0.4\% \pm 0.03\%$.
The sample distributions are not similar (the two sample Student's t-Test gives p--value = \ensuremath{3.819\times 10^{-7}}), 
therefore at least part of the measured PD of WDs from our sample is intrinsic. This seems to be confirmed by 
the spatial distribution of our WDs sample in galactic coordinates
in comparison to \cite{Berdyugin2014} measurements that is shown in Fig.~\ref{fig:gal_pol}. There are regions in the Galaxy that
show higher polarization degree, however for example DC WDs from our sample are not
located in these regions. Still, higher PD of DC WDs might be biased
by the fact that we are close to the telescope limits in terms of brightness.

\subsection{Spectral type and binarity comparison}
The comparison of PD distributions of both types of 
isolated WDs in the form of a boxplot is presented in Fig.~\ref{fig:da_vs_dc}. It can 
be seen that DC type white dwarfs on average have higher PD than DA type WDs. However, the 
sample sizes are unequal with the ratio of 10:1 in the favour of DA type WDs. For a validation a bootstrap
experiment was conducted. We drew 10000 samples of size 4 from DA type WDs
with replacement and recorded median PD value. In less than 0.01\% of all cases (1 out of 10000) the median value of PD from DA type
WDs was greater than that of DC. The median of all median PD values agrees with that of the whole isolated DA sample.

In our sample we have 21 binary systems. In Fig.~\ref{fig:iso_vs_bin}
we compare isolated DA WDs (49)
with DA+dM binaries (11)
and double degenerated systems 
(DDSs, 7) with respect to their PD. 
The samples show similar properties with comparable median PD values. Three bootstrap experiments were conducted
similarly to the previous case and they confirmed the results presented in Fig.~\ref{fig:iso_vs_bin}.
This suggests, as expected, that there is no significant contribution to the PD from 
the companion that in most cases of WDs binary systems is a red dwarf. 
There are two DA+dM outliers in Fig.~\ref{fig:iso_vs_bin},
i.e. WD1440$-$025 and WD2213+317. 
WD1440$-$025 is classified as a DA+dMe binary, while
WD2213+317 is located close to the Galactic disc ($b\simeq -20\dg$), has high 
proper motion \citep{Ivanov2008}, and at the same time, it is the faintest object in our WD sample.

\subsection{Magnetic WDs and WDs with measured magnetic field}
Magnetic fields of isolated magnetic WDs (MWDs) are in the range
between $10^{3} - 10^{9}~\rm G$, with quite well
established upper limit, but rather uncertain lower
limit \citep{Ferrario2015}. The magnetic field of white dwarfs in magnetic cataclysmic variables (MCVs) varies between $7 - 230~\rm MG$ \citep{Ferrario2015}.  A comparison of our target list with the surveys by \citet{Schmidt1995} and \citet{Putney1997} reveals that there are two highly MWDs and 20 WDs with measurable but rather low (with respect to MWDs) or not significant magnetic field values ($B \sim \rm kG$) in our sample. These 22 objects are marked with a $\dag$ in Tab.~\ref{tab:results}.

The two known isolated, highly MWDs (B > 1~MG) in our sample are WD1639+537 (GD 356) and  WD1658+440 (PG 1658+440). Both have the same DA spectral type and have similar SDSS $r$ brightness of around 15 mag. Thus taking into account only isolated WDs (53 out of 74), only 3.8\% of our sample are highly MWDs.
WD1639+537 not only has a high surface magnetic field of the order of 11.2$\pm$1.1~MG \citep{Greenstein1985}, but it is also photometrically variable. Its variability is most likely caused by the
dark spot that covers 10\% of the stellar surface \citep{Brinkworth2004}. WD1658+440 has a mean surface magnetic field on the level of 2.3$\pm$0.5~MG \citep{Liebert1983}, confirmed later by \cite{Schmidt1992b}  to be 2.3$\pm$0.2~MG. The maximum circular
spectropolarimetric values of WD1639+537 and WD1658+440 are $\pm$2-3\% 
\citep[from Fig.~1. of][]{Ferrario2015}
and 4.8\% \citep[Tab. 2 of][also later confirmed by \citealt{Schmidt1992b}, see Fig.~1 of their work with the circular polarization spectra, where maximum V values reach around $\pm$5-6\% in H$\alpha$ and H$\beta$ lines]{Liebert1983}.
Whereas their linear PD (this work) is on the level of 0.31\%$\pm$0.48\% and 0.04\%$\pm$0.52\%, respectively. However, in both cases the corrected linear PD is zero with the upper error on the level of $\sim$0.5\%. The reason for such a low linear polarization must be the same as that for the low value of circular continuum polarization ($V$=0.01$\pm$0.047\%, \citealt{Angel1981}), namely, the dilution over the broad band range of the Johnson R filter.

The fact that we have only $\sim 4$\% of isolated MWDs in our sample
is roughly consistent with results presented in the magnitude limited surveys.
As stated by \cite{Schmidt1995}, the incidence of magnetism among WDs 
is found to be 4.0\%$\pm$1.5\% for fields between $\sim 3 \times 10^4$ and $10^9$~G.
Similarly, \cite{Kepler2013} found by detectable Zeeman splittings
that around 4\% of all observed DAs have magnetic fields in the range
from  1 to 733 MG.

On the other hand, there are 20 WDs in our sample for which the magnetic field measurements were performed resulting with magnetic fields below 1~MG ($\rm{kG} < B < \rm{MG}$). 19 out of these 20 WDs
are described by \citet{Schmidt1995} and one by \citet{Putney1997}. We should note here that there is one more WD in \citet{Putney1997}  common with our list, but with no magnetic field value (see later in the paragraph for details).
From \citet{Schmidt1995} they are as follow, isolated: WD1310+583, WD1319+466, WD1344+572, WD1407+425, WD1408+323, WD1446+286, WD1553+353, WD1601+581, WD1606+422, WD1632+177, WD1637+335, WD1641+387, WD1647+375, WD1655+215, WD1713+695, WD2149+021 and in binary systems: WD1317+453, WD1659+303 and WD2126+734. The highest measured absolute value of the magnetic field strength in this group is on the level of 13.8~kG (WD1317+453),
but one should notice that in many cases 
the measurement is affected by large uncertainty. We can also compare our results with the circular spectropolarimeric survey of \citet[hereafter P97]{Putney1997}. Out of 52 WDs that she measured only two of them are in our sample, i.e. WD1257+037 (G60$-$54, $m_r = 15.6$) and 
WD1712+215 (G170$-$27, $m_r = 16.6$). The first one is of the DA type (DAQZ8 in P97), while the second is the DC type (DC7 in P97). 
Only in the first case the author was able to calculate the magnetic field based on the circularly polarized absorption lines and she obtained
$B_e = 11.2 \pm 18.0 ~\rm kG$. In case of WD1712+215 the magnetic field strength could not be estimated because it does not show 
absorption features nor circular polarization. P97 suggested an upper limit of 20~MG. Our results of 
linear PD for both WDs are $0.34\% \pm 0.36\%$ and $0.83\% \pm 0.71\%$, respectively. However, both corrected linear PD are zero with the 
upper error on the level of 0.36\% and 0.71\% for WD1257+037 and WD1712+215, respectively. While the circular polarization values are
$V_{\rm{red}}[\%] =+0.045 \pm 0.073$, $V_{\rm{blue}}[\%] =+0.092 \pm 0.156$
and $V_{\rm{red}}[\%] =+0.088 \pm 0.038$, $V_{\rm{blue}}[\%] =+0.008 \pm 0.066$ for WD1257+037 and WD1712+215, respectively (see \citealt{Putney1997} for details).

Apparently, even in case of two highly MWDs with the magnetic field on the order of a few MG we do not detect significant linear polarization degree. This is likely caused by the fact of the polarization dilution over broad band. Therefore, we conclude that the MWDs (at least those of our sample) can also be used as zero-polarimetric standards.

\section{Summary and conclusions}
\label{sec:summary}

We conclude that most of the measured WDs, being of DA and DC spectral types and in the SDSS \textit{r} magnitude range from
13 to 17,
have low PD below 1\%. Only WD1440$-$025 and WD2213+317
have PD higher than 1\%. We found out that there is a correlation between the measured PD and \textit{r} brightness, 
as well as between the $\sigma_\mathrm{PD}$ and \textit{r} brightness. Moreover, the DC type white dwarfs on average have higher PD 
(with the median PD of 0.78\%) than DA type WDs
(with the median PD of 0.36\%). The significance of the difference (p--value of the null
hypothesis) is on the level of 0.01.
Taking into account that the PD and PD uncertainties show
the dependence on the brightness the difference between those two WD types
can be attributed to the fact that the DC type WDs are fainter than DA. However, there
are only 4 DC type WDs in our sample,
so we can not state this without any doubts. 
On the other hand, there seems to be no difference between PD of isolated DA type WDs (0.36\%) 
and binary systems that include DA type WDs (0.35\%). 

Our sample constitutes a set of good candidates of 
faint linear polarimetric standard stars with SDSS \textit{r} magnitudes ranging from 13 up to 17. They are well distributed 
in the right ascension range from 13 hour up to 24 hour mostly on the Northern sky with declination form \ensuremath{-11}$\dg$ up to 78$\dg$. Moreover, we enrich the present
low linear polarization WDs standard list by a factor of five.
Reaching fainter objects with infrastructure with larger mirror introduced a serious problem, namely, the lack of faint polarization 
standards of both types, the zero-polarized and polarized ones.
The presented list of WDs addresses this need with respect to zero-polarized standard
stars and is complementary to the previous work done by \cite{Fossati2007}, which includes nine WDs in the magnitude range from 11 to 13, with only one exception of 14 mag.

Additionally, we found that for low extinction values (< 0.04) the best
model that describes the dependence of PD on E(B--V) is given by the equation:
$\rm{PD_{max, ISM}}[\%] = 0.65~E(B-V)^{0.12}$.

Even in cases of highly MWDs, as well as in the cases of low magnetic field WDs, we do not detect significant linear polarization degree. This is likely caused by the fact of the polarization dilution over broad band. Therefore, we conclude that even the MWDs of our sample can be very well used as polarimetric standards.

It will be very useful to perform deeper and longer observations
on the DC type WDs to obtain measurements with higher accuracy
and to expand the test sample. It is worth mentioning that Gaia satellite
will discover around 100,000 WDs. Assuming the same ratio of around 300/23,000 (WDs brighter
than 17 mag visible from the Skinakas Observatory to the total number of known WDs),
there will be around 1,300 Gaia WDs brighter than 17 mag visible at Skinakas,
therefore we will be able to continue our study in the future on much bigger sample.

\section*{Acknowledgements}

This work has been supported by Polish National Science Centre grant
DEC-2011/03/D/ST9/00656 (AS, KK, M\.Z). This research was partly supported by 
the EU COST Action MP1104 "Polarization as a tool to study the solar system and
beyond" within STSM projects: COST-STSM-MP1104-14064, COST-STSM-MP1104-16823
COST-STSM-MP1104-14070 and COST-STSM-MP1104-16821.
DB acknowledges support from the St. Petersburg Univ. research grant 6.38.335.2015.
RoboPol is a collaboration involving the University of Crete, the
Foundation for Research and Technology - Hellas, the California
Institute of Technology, the Max-Planck Institute for Radioastronomy,
the Nicolaus Copernicus University, and the Inter-University
Centre for Astronomy and Astrophysics. This work was partially
supported by the "RoboPol" project, which is co-funded by the European Social
Fund  (ESF) and Greek National Resources, and by the European Comission Seventh Framework
Programme (FP7) through grants PCIG10-GA-2011-304001 "JetPop" and PIRSES-GA-2012-31578 "EuroCal". Data analysis and figures were partly prepared using R \citep{rcite}.

\bibliographystyle{mnras}
\bibliography{zejmo}


\label{lastpage}
\end{document}